\documentclass[prc,aps,float,nofootinbib,superscriptaddress,twocolumn]{revtex4-1}
\usepackage{amssymb}
\usepackage{amsmath}
\usepackage{amsfonts}
\usepackage{epsfig}
\usepackage{bbm}
\usepackage{comment}
\usepackage{multirow}
\usepackage{longtable}
\usepackage{array}
\usepackage{subfigure}
\usepackage{lipsum}
\usepackage{array}
\usepackage[usenames,dvipsnames]{color}
\usepackage[colorlinks]{hyperref}
\def\ema{$N\!N\!$+$3N(400)$}
\def\lnl{$N\!N\!$+$3N\text{(lnl)}$}
\def\sat{NNLO$_{\text{sat}}$}
\def\gaz{$2N$+$3N(500)_{c_D\!0.83}$}
\def\heb{$1.8/2.0$ (EM)}

\allowdisplaybreaks

\begin{document}
\title{Novel chiral Hamiltonian and observables in light and medium-mass nuclei}
\author{V. Som\`a}
\email{vittorio.soma@cea.fr} 
\affiliation{IRFU, CEA, Universit\'e Paris-Saclay, 91191 Gif-sur-Yvette, France} 
\author{P. Navr\'atil}
\email{navratil@triumf.ca}
\affiliation{TRIUMF, 4004 Westbrook Mall, Vancouver, British Columbia, V6T 2A3, Canada}
\author{F. Raimondi}
\email{francesco.raimondi@cea.fr}
\affiliation{ESNT, IRFU, CEA, Universit\'e Paris-Saclay, 91191 Gif-sur-Yvette, France} 
\affiliation{Department of Physics, University of Surrey, Guildford GU2 7XH, United Kingdom}
\author{C. Barbieri}
\email{c.barbieri@surrey.ac.uk}
\affiliation{Department of Physics, University of Surrey, Guildford GU2 7XH, United Kingdom}
\author{T. Duguet}
\email{thomas.duguet@cea.fr} 
\affiliation{IRFU, CEA, Universit\'e Paris-Saclay, 91191 Gif-sur-Yvette, France} 
\affiliation{KU Leuven, Instituut voor Kern- en Stralingsfysica, 3001 Leuven, Belgium}

\date{\today}

\begin{abstract} 
\begin{description}
\item[Background]  
Recent advances in nuclear structure theory have led to the availability of several complementary \textit{ab initio} many-body techniques applicable to light and medium-mass nuclei as well as nuclear matter. 
After successful benchmarks of different approaches, the focus is moving to the development of improved models of nuclear Hamiltonians, currently representing the largest source of uncertainty in \textit{ab initio} calculations of nuclear systems.
In particular, none of the existing two- plus three-body interactions is capable of satisfactorily reproducing all the observables of interest in medium-mass nuclei.
\item[Purpose]  
A novel parameterisation of a Hamiltonian based on chiral effective field theory is introduced.
Specifically, three-nucleon operators at next-to-next-to-leading order are combined with an existing (and successful) two-body interaction containing terms up to next-to-next-to-next-to-leading order.
The resulting potential is labelled \lnl.
The objective of the present work is to investigate the performance of this new Hamiltonian across light and medium-mass nuclei.
\item[Methods] 
Binding energies, nuclear radii and excitation spectra are computed using state-of-the-art no-core shell model and self-consistent Green's function approaches.
Calculations with \lnl{} are compared to two other representative Hamiltonians currently in use, namely \sat{} and the older \ema.
\item[Results]
Overall, the performance of the novel \lnl{} interaction is very encouraging. 
In light nuclei, total energies are generally in good agreement with experimental data.
Known spectra are also well reproduced with a few notable exceptions.
The good description of ground-state energies carries on to heavier nuclei, all the way from oxygen to nickel isotopes.
Except for those involving excitation processes across the $N=20$ gap, which is overestimated by the new interaction, spectra are of very good quality, in general superior to those obtained with \sat{}.
Although largely improving on \ema{} results, charge radii calculated with \lnl{} still underestimate experimental values, as opposed to the ones computed with \sat{} that successfully reproduce available data on nickel.
\item[Conclusions] 
The new two- plus three-nucleon Hamiltonian introduced in the present work represents a promising alternative to existing nuclear interactions. 
In particular, it has the favourable features of (i) being adjusted solely on $A=2,3,4$ systems, thus complying with the \textit{ab initio} strategy, (ii) yielding an excellent reproduction of experimental energies all the way from light to medium-heavy nuclei and (iii) behaving well under similarity renormalisation group transformations, with negligible four-nucleon forces being induced, thus allowing large-scale calculations up to medium-heavy systems.
The problem of the underestimation of nuclear radii persists and will necessitate novel developments.
\end{description}
\end{abstract}
\maketitle

\section{Introduction}
\label{sec_introduction}

In the last decade, advances in many-body approaches and inter-nucleon interactions have enabled significant progress in \textit{ab initio} calculations of nuclear systems. 
At present, several complementary methods to solve the (time-independent) many-body Schr\"odinger equation are available, tailored to either light systems~\cite{Navratil16, Carlson15}, medium-mass nuclei~\cite{Dickhoff04, Lahde14, Hagen14, Hergert16, Stroberg17, Tichai18a} or extended nuclear matter~\cite{Carbone16, Lynn16, Drischler19}. 
New developments, which promise to extend (most of) these methods to higher accuracy and/or heavy nuclei, are being currently proposed~\cite{Tichai2019Tdec, Tichai:2019ksh}.

Over the past few years, benchmarks calculations have allowed assessment of the systematic errors associated with both the use of a necessarily finite-dimensional Hilbert space and the truncation of the many-body expansion at play in each of the formalisms of interest. 
In state-of-the-art implementations, these errors add up to at most $5\%$, much less than the uncertainty attributable to the input nuclear Hamiltonian~\cite{Hergert14, Hebeler15, Lapoux16, Duguet17a, Leistenschneider18}. 
As a result, \textit{ab initio} calculations have also acquired the role of diagnostic tools as the focus of the community is shifting towards developing improved models of nuclear interactions.

The large majority of these developments currently takes place in the context of chiral effective field theory ($\chi$EFT)~\cite{Epelbaum2009RMP, Machleidt11} based on Weinberg's power counting (WPC)~\cite{Weinberg90, weinberg91, Ordonez92}. 
Building on a low-energy expansion with nucleons and pions as explicit degrees of freedom, $\chi$EFT provides a framework in which two- and many-nucleon interactions can be systematically derived, in principle with uncertainties associated with each order of the expansion. 
This approach to $\chi$EFT suffers from issues related to its non renormalizability and different strategies have been proposed to resolve this problem, requiring the use of a different power counting and a mixture of non-perturbative and perturbative calculations~\cite{Hammer2019EFT}. However, applications of these procedures are still in their infancy stage.
From the practical point of view, the standard implementations of $\chi$EFT interactions are much more advanced and therefore in better position to provide useful predictions.
The present study follows the conventional approach of seeking fully non-perturbative solutions of the many-body Schr\"odinger equation and focuses on the performance of \textit{ab initio} calculations for isotopes up to medium masses.

Within this framework, one of the first successful nuclear Hamiltonians combined a next-to-next-to-next-to-leading order (N$^{3}$LO) two-nucleon ($2N$) force~\cite{Entem03} with a local N$^{2}$LO three-nucleon ($3N$) interaction~\cite{Navratil07}, whose associated cutoff was subsequently reduced to 400 MeV/c to optimise its behaviour under similarity renormalization group (SRG) transformations~\cite{Roth12}. 
This Hamiltonian, labelled \ema, has constituted a standard for many early applications of \textit{ab initio} techniques in light as well as medium-mass nuclei. 
As systematic calculations beyond the light sector became available, deficiencies associated to this interaction emerged.  
In particular, it was shown to lead to an underestimation of nuclear radii and a substantial overbinding, i.e. an overestimation of total binding energies in medium- and heavy-mass nuclei~\cite{Soma14b, Binder14, Lapoux16}. Recently, this behaviour was related to the poor description of saturation properties of symmetric nuclear matter~\cite{Simonis17, Drischler19}.

While the initial success of \ema{}  represented an important breakthrough, its poor performance for isotopes above the oxygens prompted new work aimed at improving predictions of nuclear saturation. A successful route was put forward by Ekstr\"om and collaborators in Ref.~\cite{Ekstrom15}. There, in contrast to the standard strategy of constraining the Hamiltonian only on few-nucleon data, the binding energies and charge radii of nuclei up to $A=25$ were employed in the simultaneous fit of the low-energy constants (LECs) that enter the $2N$ and $3N$ interactions.  The resulting next-to-next-to-leading order Hamiltonian, labelled \sat, provides a much improved description of charge radii and fair saturation properties of extended matter. However, this is obtained at the price of deteriorating certain properties of two- and few-body systems~\cite{Ekstrom15, Lapoux16}. Furthermore, mild underbinding is observed in medium-mass isotopes.
Another family of Hamiltonians that have proven to perform well for \textit{either} binding energies \textit{or} radii~\cite{Simonis17} was introduced by Hebeler and collaborators in Ref.~\cite{Hebeler11}. One drawback of these interactions is that $2N$ and $3N$ sectors are not consistently evolved under SRG, which introduces an additional source of uncertainty.

In the meantime, two problematic features of \ema{} were identified, both in the $3N$ sector. 
The first one concerns the LEC $c_D$ entering both the $3N$ contact operator and the contact axial current. 
This constant was originally fitted, via its contribution to the axial current in $\beta$-decay processes, to the triton half-life~\cite{Gazit09}. Recently, an error in the derivation of the axial current was pointed out~\cite{Schiavilla18, Gazit19}, which raises questions about the pertinence of the previous fit, as well as about subsequent many-body calculations that employed this interaction. 
The second point relates to possible artefacts introduced when certain types of regulator functions are used.
In principle, any function suppressing high momentum modes could be employed.
In practice, however, it was shown~\cite{Hagen2014InfNM,Dyhdalo16,Lynn2017QMClight,Lu2019binding} that local regulators, as employed for $3N$ operators in the standard \ema{} implementation, are likely to cut off regions of the phase space that are instead relevant.
The success of \sat was also attributed in part to the use of nonlocal regulators in the $3N$ interaction.

To overcome the above difficulties, a variant of the \ema{} interaction that remedies the contact axial current fit of the LECs and employs both local and nonlocal (lnl) $3N$ regulators is introduced and tested in the present work. 
This new Hamiltonian is labelled \lnl. 
Low-energy constants of  $2N$ and $3N$ operators are fitted to properties of $A=2,3,4$ nuclei, for which an excellent description is maintained including the triton half-life once the correct contact axial current is employed. 
Properties of light nuclei are investigated by means of no-core shell model~(NCSM)~\cite{Barrett2013} calculations.
Encouragingly, results are generally in good agreement with experiment.
In order to assess its quality in medium-mass nuclei, many-body calculations are performed within the self-consistent Green's function approach, in both its closed-shell (i.e. Dyson)~\cite{Cipollone13} and open-shell (i.e. Gorkov)~\cite{Soma11a} versions. 
Three representative isotopic chains, namely oxygen, calcium and nickel, are addressed.
Total binding energies, two-neutron separation energies, charge radii and density distributions, as well as one-nucleon addition and removal spectra are systematically analysed and compared with calculations performed with other Hamiltonians. 
Overall, for what concerns ground- and excited-state energies, the performance of \lnl{} is very satisfactory, in line with \sat{} or even superior in certain regions.  
Only for radii (and densities) does the picture change, with \lnl{} improving with respect to \ema{} but still underestimating experimental data and \sat{} results. Possible sources of such disagreement are discussed.

The paper is organised as follows. 
The \hbox{$2N+3N$} Hamiltonians used in the present study are described in Sec.~\ref{sec_H}, with particular emphasis on the novel \lnl{} interaction.
Section~\ref{sec_light} discusses the performance of the new Hamiltonian in light systems, focusing on total energies and excitation spectra across $s$-~and $p$-shell nuclei.
Results for medium-mass nuclei are presented in Sec.~\ref{sec_medium}.
First, the employed many-body method is introduced (Sec.~\ref{sec_GF}) and convergence of the calculations with respect to model space (Sec.~\ref{sec_MSconv}) and many-body truncations (Sec.~\ref{sec_MBconv}) are discussed.
Next, results for several observables are systematically studied: ground-state energies (Sec.~\ref{sec_GS}), charge radii and distributions (Sec.~\ref{sec_radii}), and spectra of odd-$A$ nuclei (Sec.~\ref{sec_spectra}).
Concluding remarks are presented in Sec.~\ref{sec_discussion}.

\section{Hamiltonians}
\label{sec_H}

Three different Hamiltonians are employed in this work. 
The first one, labelled \ema, is based on the chiral N$^3$LO nucleon-nucleon potential from Entem and Machleidt~\cite{Entem03,Machleidt11} combined with the chiral N$^2$LO $3N$ interaction with a local regulator~\cite{Navratil07}. The $2N$ interaction of Ref.~\cite{Entem03} was built with a cutoff of 500 MeV/c, however, a 400 MeV/c regulator was used for the $3N$ sector~\cite{Roth12}.
This Hamiltonian has been employed extensively in calculations of $p$- and $sd$-shell nuclei and describes well the binding energy of oxygen, nitrogen and fluorine isotopes~\cite{Hergert13,Cipollone13}. 
Nevertheless, as also shown in this paper, it leads to overbinding in medium-mass nuclei starting in the calcium chain and underpredicts radii even for O isotopes~\cite{Binder14, Soma14b, Lapoux16}. 
In this Hamiltonian, the $3N$ low-energy constant $c_D$ had been set to -0.2 based on a $^3$H half-life fit in Ref.~\cite{Gazit09}.
However, recently it was pointed out that an error was present in the relationship between the two-body axial current LEC $d_R$ and the $c_D$ (specifically, a missing factor of -1/4) and that the $c_D$ that actually fits the $^3$H half-life is 0.83(24)~\cite{Gazit19}.
The $c_E=0.098$ LEC was determined by fitting the $^4$He binding energy~\cite{Roth12} after the cutoff was reduced to 400~MeV/c from the original 500~MeV/c~\cite{Gazit09} to mitigate four-body terms induced by the SRG evolution~\cite{Roth12}.

With the main goal of improving the description of radii in medium-mass nuclei, a new chiral Hamiltonian with terms up to N$^2$LO was developed in Ref.~\cite{Ekstrom15}.
It is characterised by a simultaneous fit of $2N$ and $3N$ LECs that does not rely solely on two-nucleon and $A$=3,4 data, but also on binding energies of $^{14}$C and $^{16,22,24,25}$O as well as charge radii of $^{14}$C and $^{16}$O. 
The resulting interaction, named \sat, successfully describes the saturation of infinite nuclear matter~\cite{Ekstrom15}, the proton radius of $^{48}$Ca~\cite{Ha16} and the nuclear radii of neutron-rich carbon isotopes as well as other medium mass nuclei \cite{Ka16, Lapoux16}. 
It also performed well in several other applications, including the description of the parity inversion in $^{11}$Be~\cite{Calci2016}, the $^{10}$C($p,p$)$^{10}$C elastic scattering~\cite{Kumar2017}, electron scattering~\cite{Rocco2015escat,Barbieri2019Ar40}, giant dipole resonances~\cite{Raimondi19}, and the derivation of microscopic optical potentials~\cite{Idini2019OpPot}. 
Unlike the \ema{} interaction, \sat{} employs a non-local regulator.

\begin{table}[t]
\begin{center}
\begin{ruledtabular}
\begin{tabular}{l|cccccc}
$^4$He: &$\langle H \rangle$ & $\langle c_1 \rangle$ & $\langle c_3 \rangle$ & $\langle c_4 \rangle$ & $\langle c_D \rangle$ & $\langle c_E \rangle $ \\
\hline
\ema{}                   & -28.28  & -0.06  &  1.27   & -3.93 & -0.28 & -0.66  \\
\sat{}       & -28.43  & -0.24  & -0.73  & -3.76 &   1.39 &  0.42  \\
\lnl                      & -28.25  & -0.18  & -1.36  & -3.27 &   0.74  &  0.43  \\                       
\gaz & -28.36  & -0.26  & -1.50  & -3.79 &   0.78  &  0.30
\end{tabular}
\end{ruledtabular}
\caption{$^4$He ground-state energies and mean values of the five chiral $3N$ N$^2$LO terms (in MeV) for the \ema~\cite{Roth12}, \sat~\cite{Ekstrom15}, \lnl{} (present work), and \gaz{} \cite{Gazit19} Hamiltonians. 
All interactions are bare, i.e. not evolved via SRG techniques.
The experimental $^4$He ground-state energy is -28.29 MeV.}
\label{tab4He}
\end{center} 
\end{table}
Motivated by the success of \sat, the objective of the present work is to amend the original \ema{} interaction, and in particular its $3N$ part.
While the latter has been shown to be problematic, its $2N$ part is instead believed to perform relatively well and thus is kept unchanged.
Being based on the N$^3$LO potential, which provides a better description of nucleon-nucleon data compared to the lower-order \sat, it guarantees superior features in light systems, e.g. a better reproduction of spectroscopy of natural parity states in $p$- and light $sd$-shell nuclei. 
A comparison of the different Hamiltonians in the calculation of the ground-state energy of $^4$He is shown in Table~\ref{tab4He}, where mean values of the five $3N$ N$^2$LO terms are displayed. 
Curiously, for \ema{} and \sat, $\langle c_3 \rangle$, $\langle c_D \rangle$, and $\langle c_E \rangle$ terms contribute with opposite sign. 
This is particularly disturbing for the three-nucleon contact term,~$c_E$.
In fact, one might argue that this could be at the origin of the severe overbinding generated by \ema{} in medium-mass nuclei.
Consequently, here the $c_D$ and $c_E$ LECs are changed to 0.7 and -0.06, respectively, to get $^4$He results more in line with the ones of \sat. 
In addition, the regulator of the $3N$ interaction was modified by introducing a non-local regulator of the same type as that used in the \sat{} on top of the local one employed in \ema{}.
For technical reasons, a completeness in the three-nucleon antisymmetrized harmonic-oscillator (HO) basis was applied to evaluate the matrix elements of the new $3N$ interaction. 
The non-local regulator was set to 500 MeV/c to be consistent with the cutoff of the N$^3$LO $2N$ interaction and the local regulator was increased from 400 MeV/c to 650 MeV/c to achieve a larger binding in A=3,4 systems in agreement with experiment. 
With these LECs, the new interaction reproduces very reasonably experimental ground-state energies of $^3$H, $^3$He and $^4$He, as well as the $^3$H half-life. 
\begin{table}[b]
\begin{center}
\begin{ruledtabular}
\begin{tabular}{l|ccccc}
              &  $c_1$  & $c_3$  &  $c_4$  &  $c_D$  &  $c_E$ \\
\hline
\ema      & -0.81    &  -3.20     & 5.40    & -0.20     & 0.098  \\
\sat        & -1.122  &  -3.925  & 3.766  &   0.817 &  -0.040  \\
\lnl         & -0.81    &  -3.20     & 5.40   &   0.70    &  -0.06  \\                       
\gaz       & -0.81    &  -3.20     & 5.40    &   0.83    &  -0.052
\end{tabular}
\end{ruledtabular}
\caption{Values of the five LECs ($c_1$, $c_3$, $c_4$ in GeV$^{-1}$) that define the $3N$ force at N$^2$LO for the \ema~\cite{Roth12}, \sat~\cite{Ekstrom15}, \lnl{} (present work), and \gaz~\cite{Gazit19} Hamiltonians.   The  $c_1$,  $c_3$ and $c_4$ constants enter both the $2N$ and the $3N$ sectors of the Hamiltonian, they are equal for \ema, \lnl{} and \gaz{}~since these share the same bare $2N$ interaction from Ref.~\cite{Entem03}. }
\label{tabLEC}
\end{center} 
\end{table}

\begin{figure*}
\centering
\includegraphics[width=5.8cm]{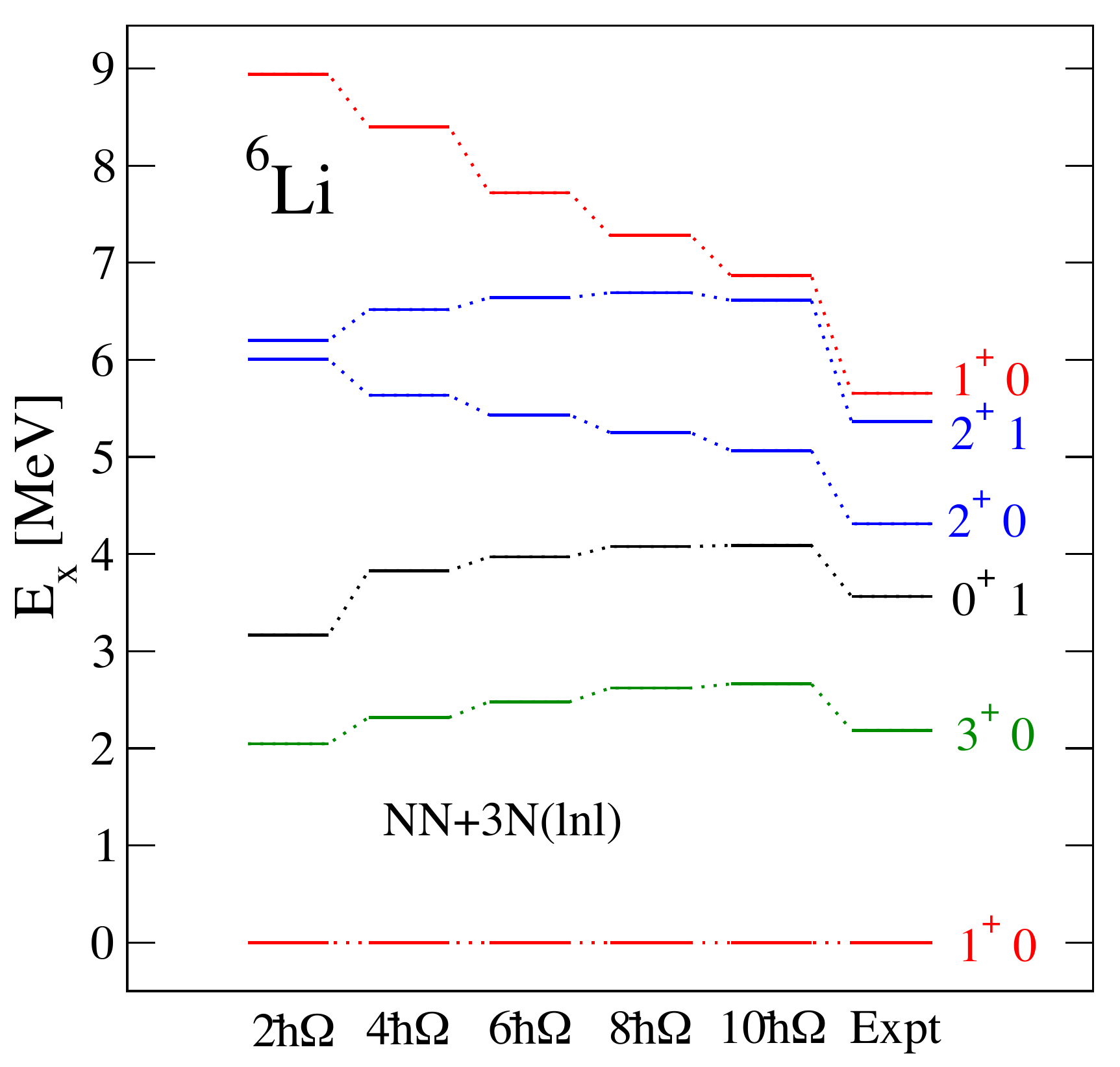}
\includegraphics[width=5.8cm]{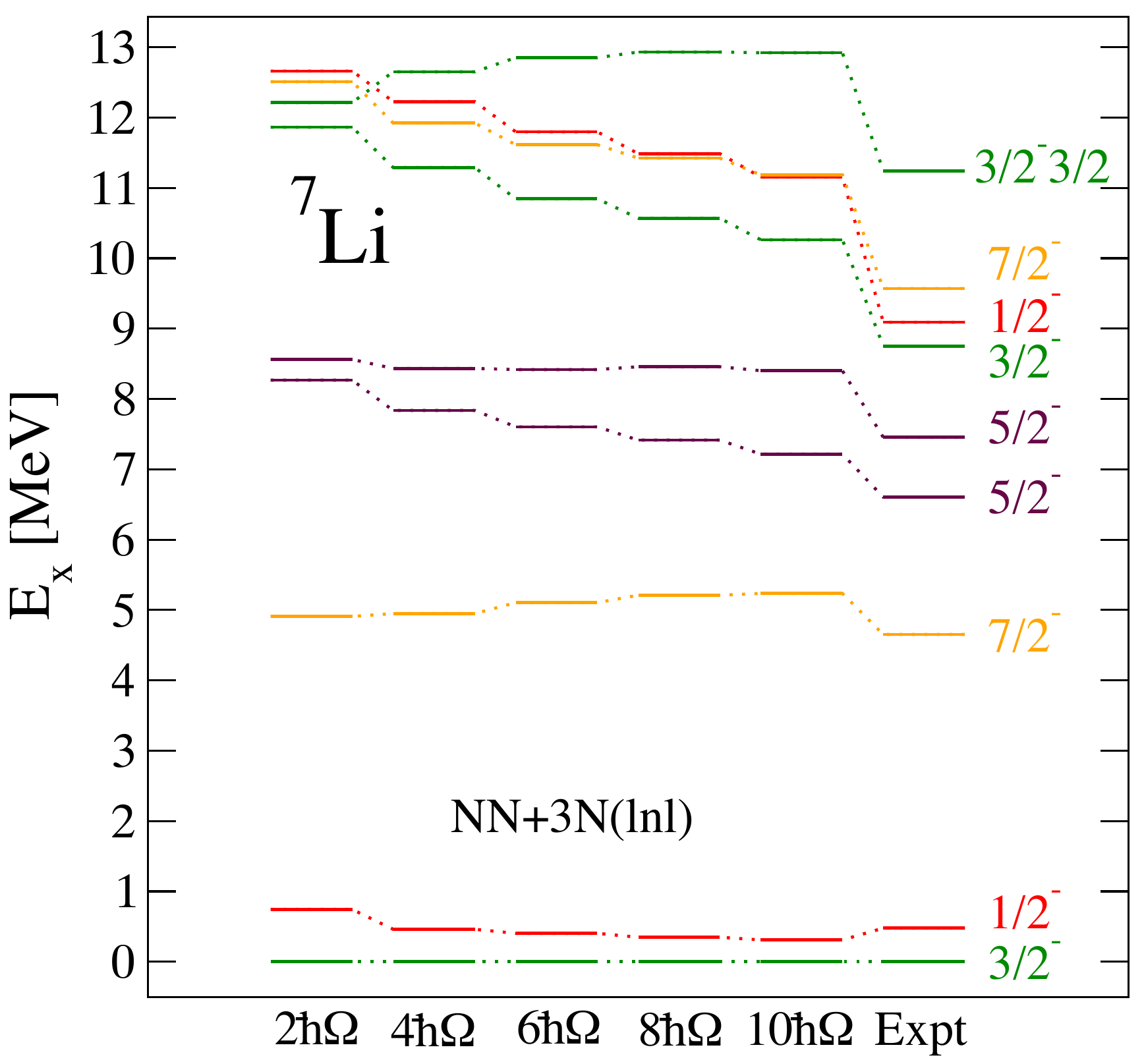}
\includegraphics[width=5.8cm]{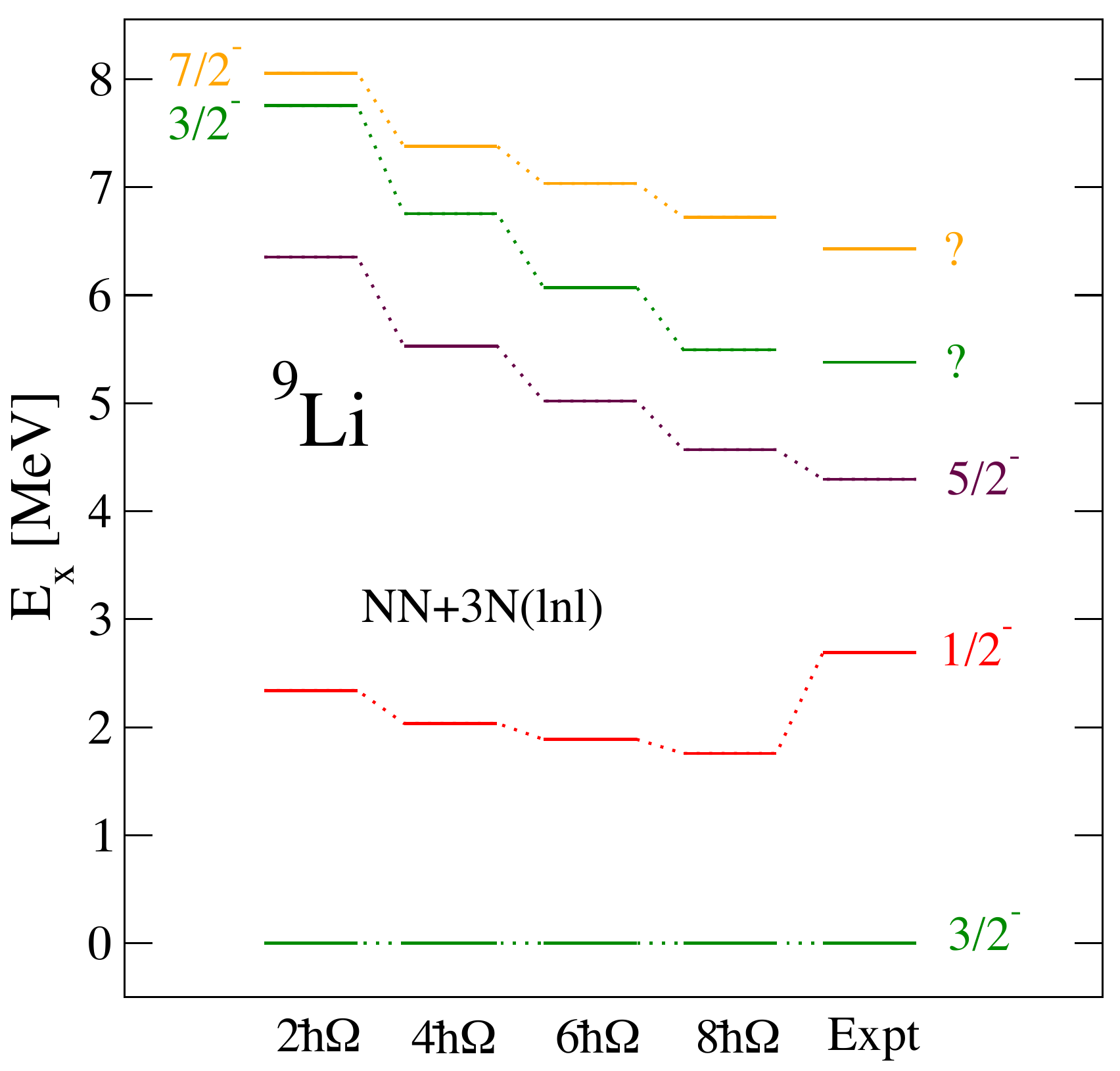}
\caption{Excited-state energies of $^{6,7,9}$Li isotopes. 
NCSM calculations with the \lnl{} Hamiltonian are compared to available experimental data.
The dependence on the NCSM basis size for $N_{\rm max}{=}2-10$ ($N_{\rm max}{=}2-8$ for $^9$Li) is shown. 
SRG evolution with $\lambda{=}2$ fm$^{-1}$ and HO frequency of $\hbar\Omega{=}20$ MeV were used.
Question marks indicate experimental levels with unassigned spin-parity values.}
\label{fig_Li}
\end{figure*}
The performance of this new interaction, named \lnl, in $^4$He is shown in Table~\ref{tab4He}. 
One observes that all $3N$ terms contribute with the same sign as in \sat{} and the $\langle c_E \rangle$ is about the same. 
The \lnl{} interaction was already applied in Ref.~\cite{Leistenschneider18} to the description of binding energies of neutron-rich titanium isotopes and neighbouring isotopic chains.
Later on, it was employed in the calculation of beta decays of selected light and medium mass nuclei~\cite{Gysbers19}, where it was denoted NN-N$^3$LO+3N$_{\rm lnl}$.
Very recently, it was also used in a study of quasifree nucleon knockout from $^{54}$Ca~\cite{Chen19}.
Its performance in these preliminary applications was very promising, which is confirmed in the present systematic and extensive study. 
It turns out, however, that the good quality of this new interaction (and of \sat) in medium-mass nuclei is not just a consequence of a particular choice of the LECs $c_D$ and/or $c_E$. 
In Table~\ref{tab4He}, results of the corrected \gaz{} interaction from Ref.~\cite{Gazit19} are also shown.
As compared to \lnl, this Hamiltonian has the same $c_1,c_3,c_4$ LECs (i.e. those from the $2N$ N$^3$LO interaction~\cite{Entem03} also used in \ema) and almost identical $c_D=0.83$ and $c_E=-0.052$, but it employs a local regulator as the \ema{} although with a 500 MeV/c cutoff. 
One can see that it gives very similar results in $^4$He as the \lnl. 
However, it severely overbinds medium-mass nuclei in many-body perturbation theory calculations~\cite{Takayuki2019}. 
Consequently, the choice of the regulator appears crucial for the correct description beyond light nuclei and perhaps a hint of the reason of the superior performance of \lnl{} manifests in the $^4$He results by the absolute values of $\langle c_3 \rangle$ and $\langle c_4 \rangle$ reduced while the $\langle c_E \rangle $ contribution enhanced compared to \gaz{}.
Table~\ref{tabLEC} summarises the LECs used to construct the 3N interactions of all the  above Hamiltonians.

\section{Light nuclei} 
\label{sec_light}

\begin{figure*}
\centering
\includegraphics[width=5.8cm]{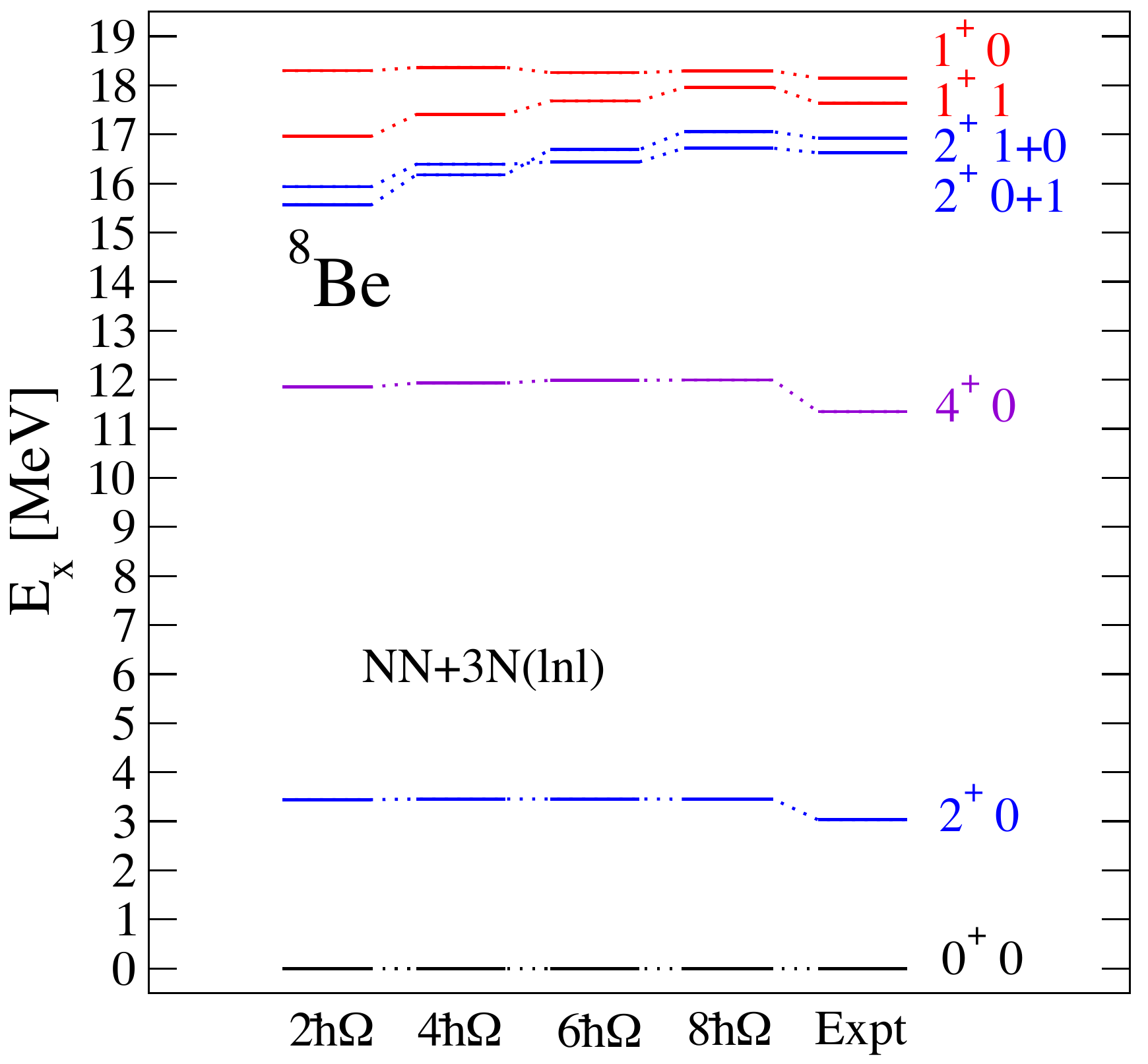}
\includegraphics[width=5.8cm]{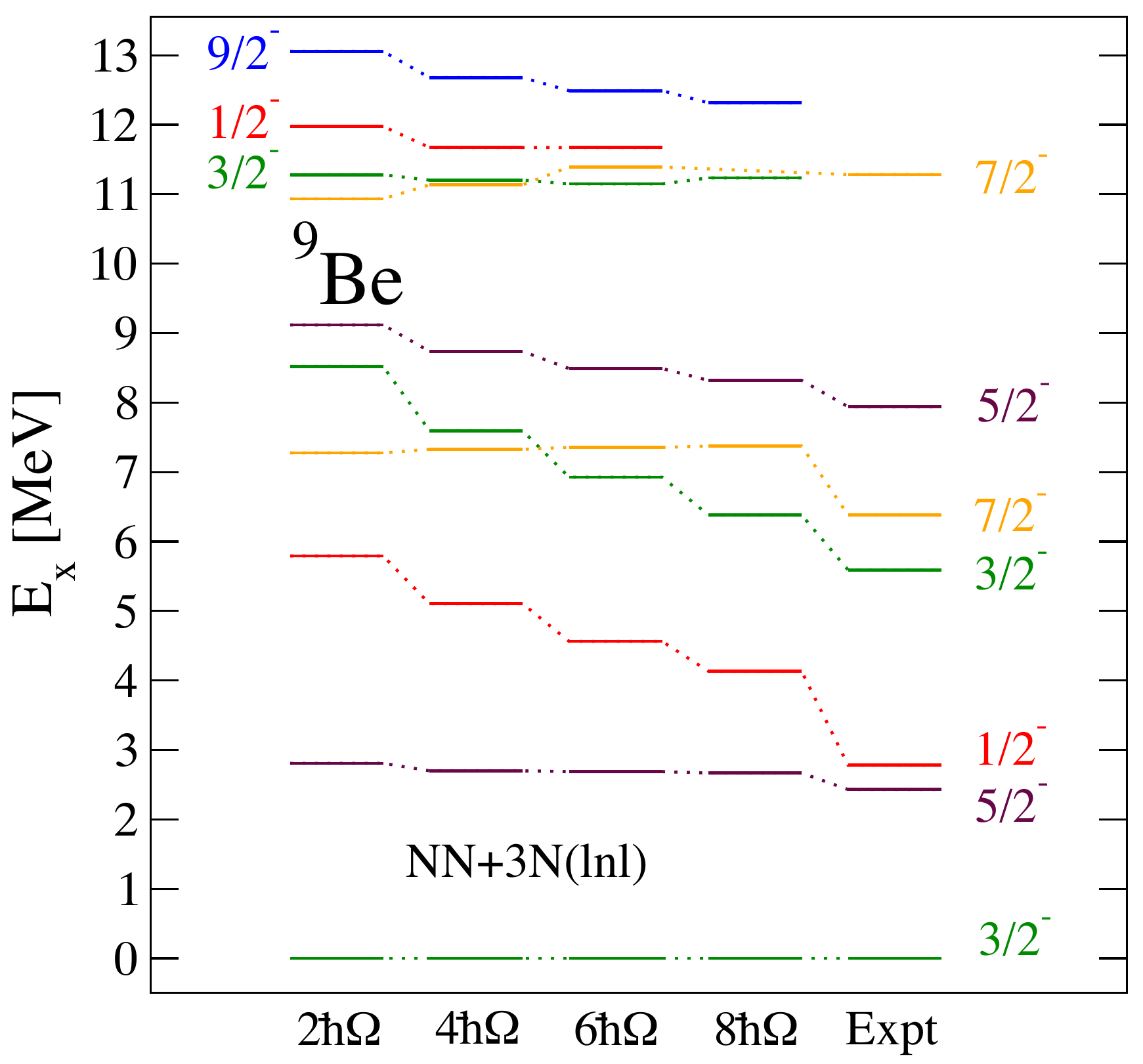}
\includegraphics[width=5.8cm]{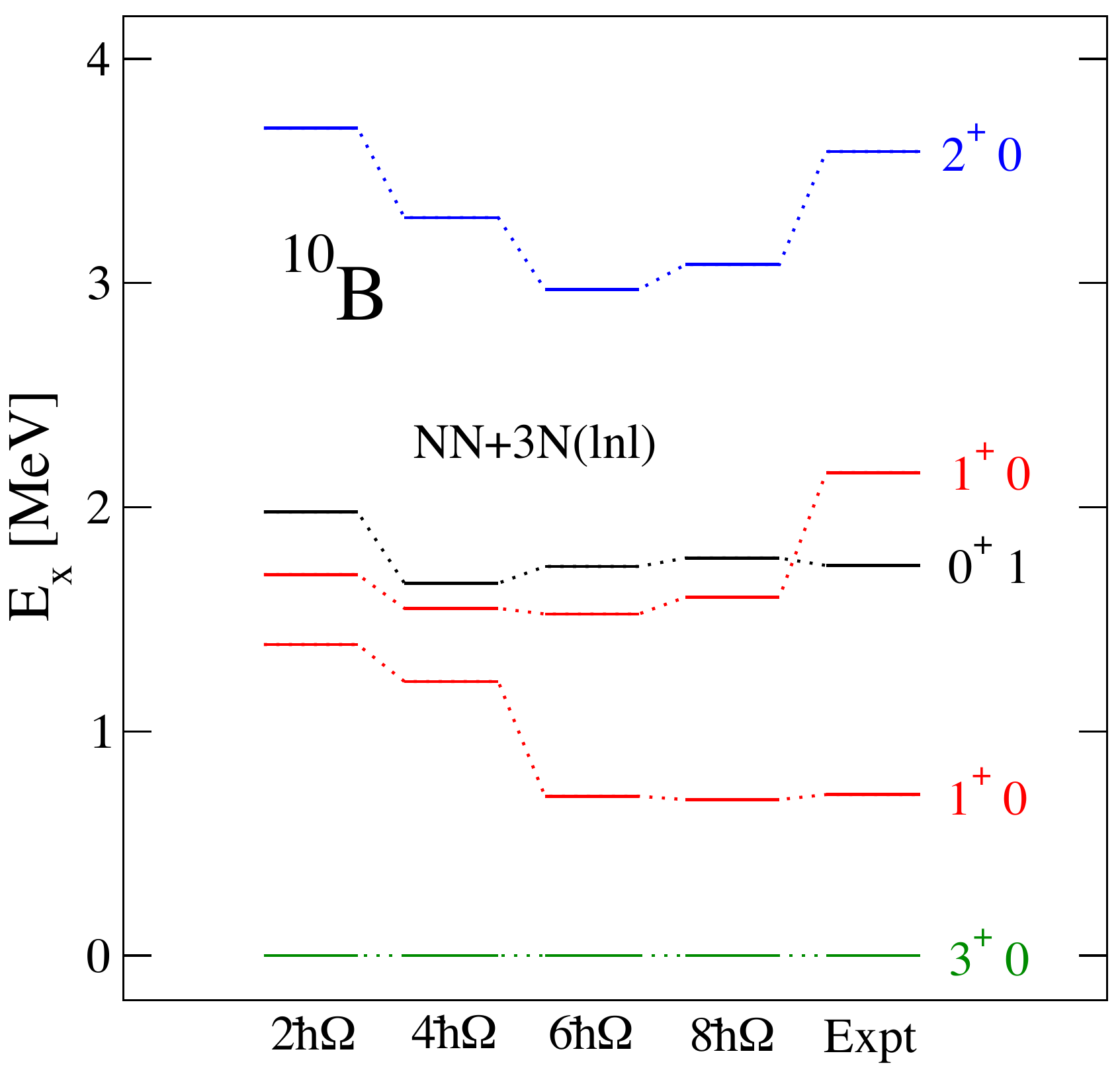}
\caption{The same as in Fig.~\ref{fig_Li} for $^{8,9}$Be and $^{10}$B. Basis sizes $N_{\rm max}{=}2-8$ are displayed.}
\label{fig_BeB}
\vskip 0.5cm
\end{figure*}
\begin{figure*}
\centering
\includegraphics[width=5.8cm]{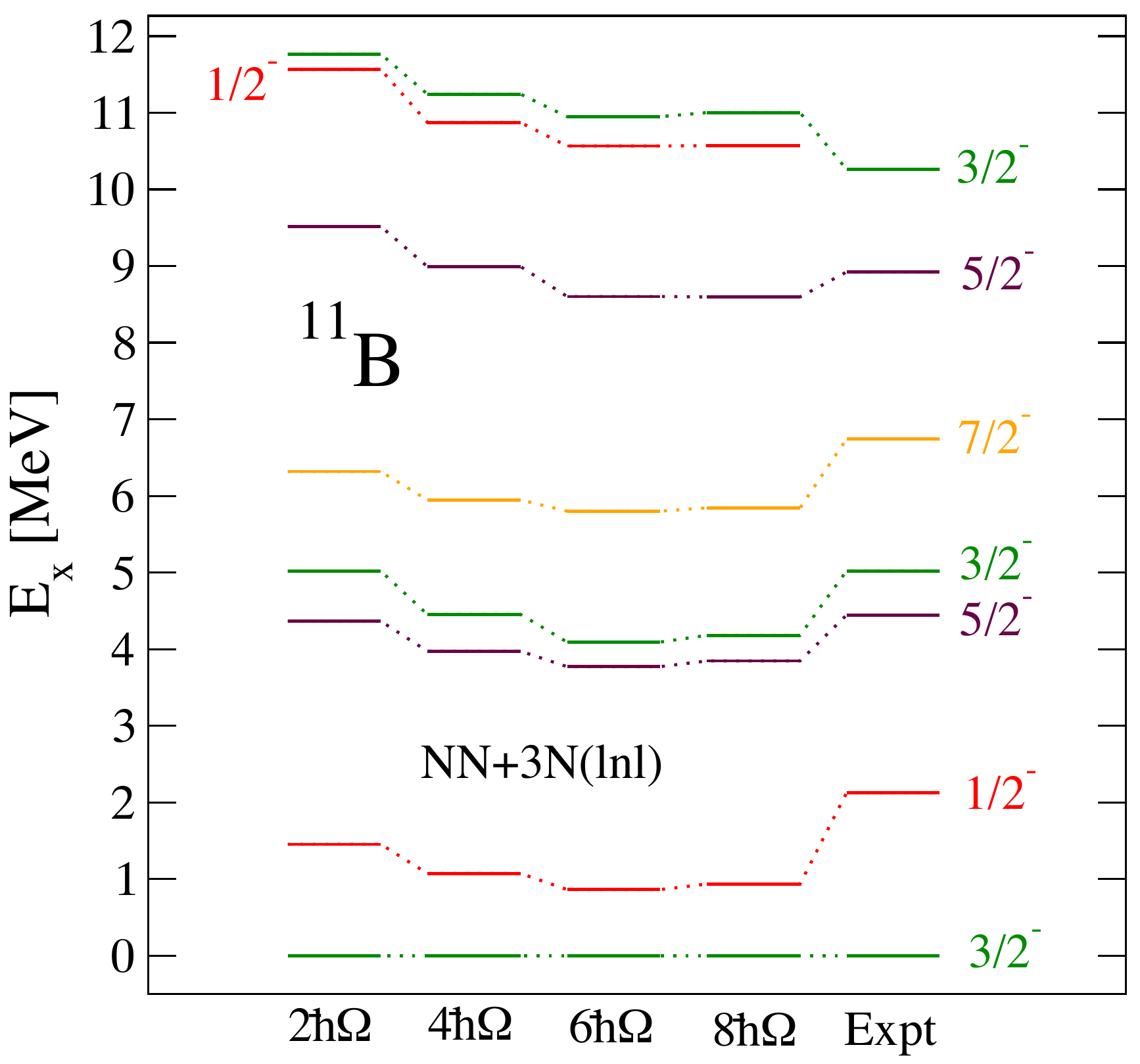}
\includegraphics[width=5.8cm]{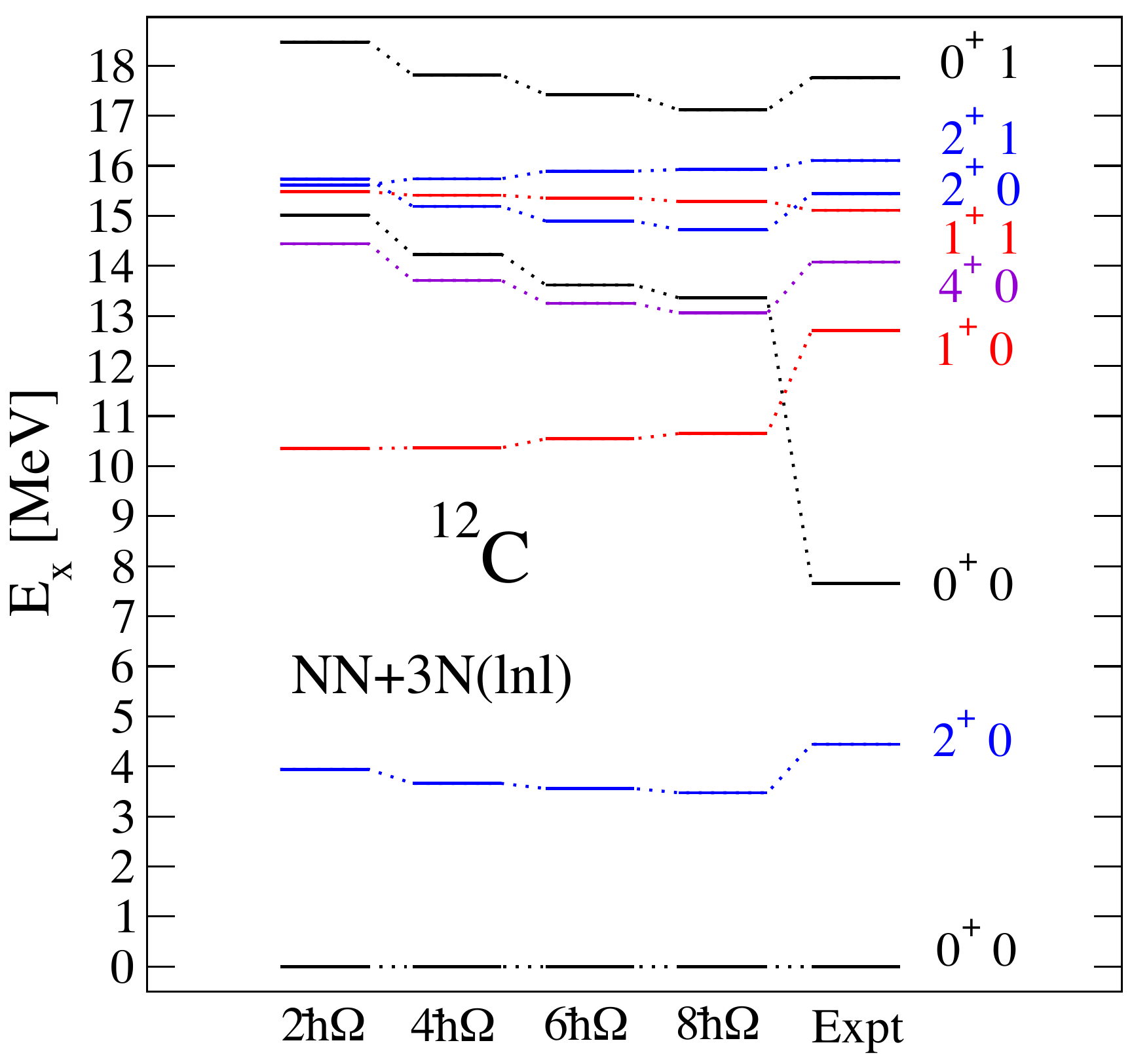}
\includegraphics[width=5.8cm]{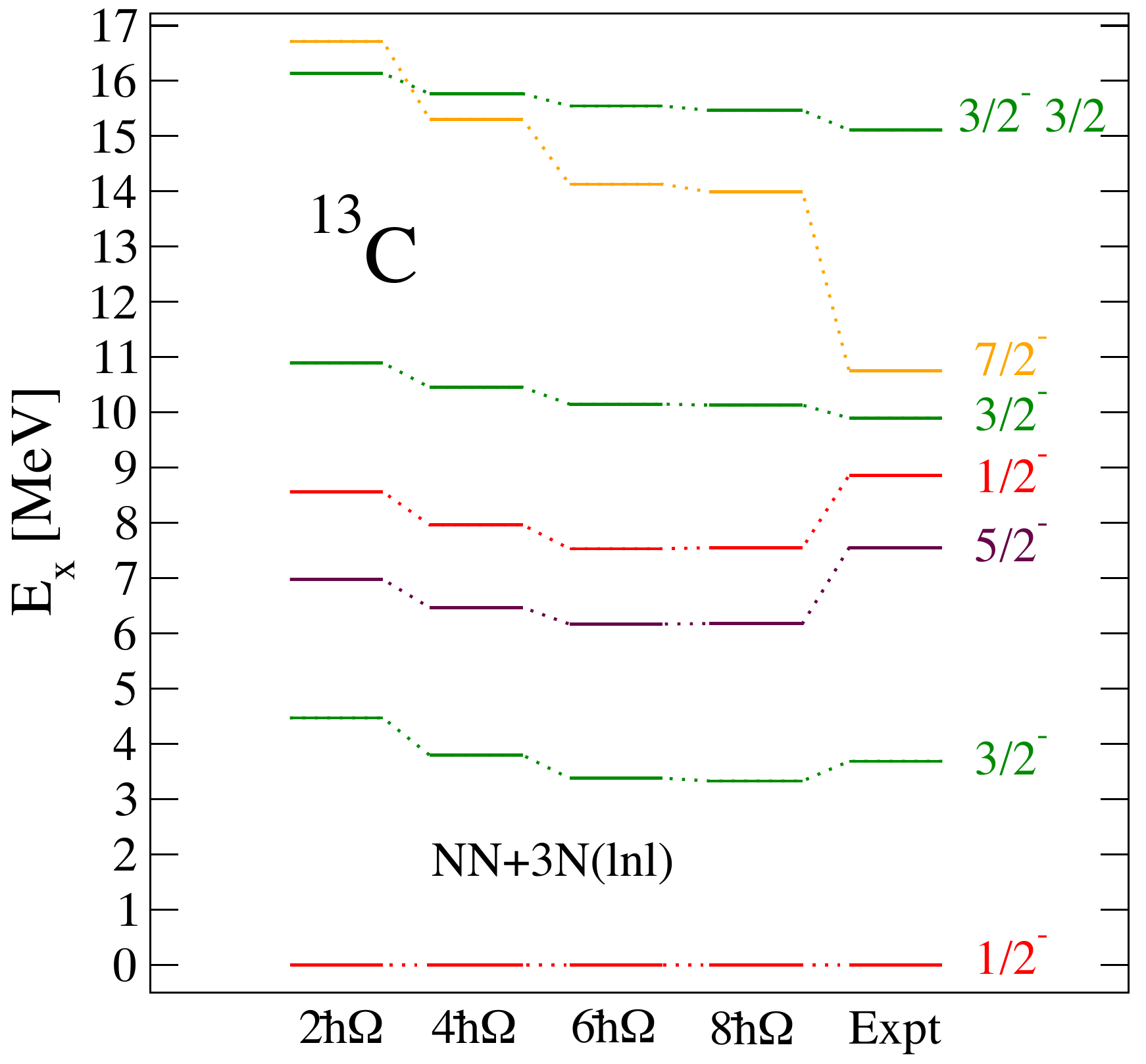}
\caption{The same as in Fig.~\ref{fig_Li} for $^{11}$B and $^{12,13}$C. 
Basis sizes $N_{\rm max}{=}2-8$ are displayed. 
The importance-truncated NCSM~\cite{Roth2007,Roth2009} was used in the $N_{\rm max}{=}8$ space for carbon isotopes.}
\label{fig_BC}
\vskip 0.5cm
\end{figure*}
\begin{figure}[t]
\centering
\includegraphics[width=8.5cm]{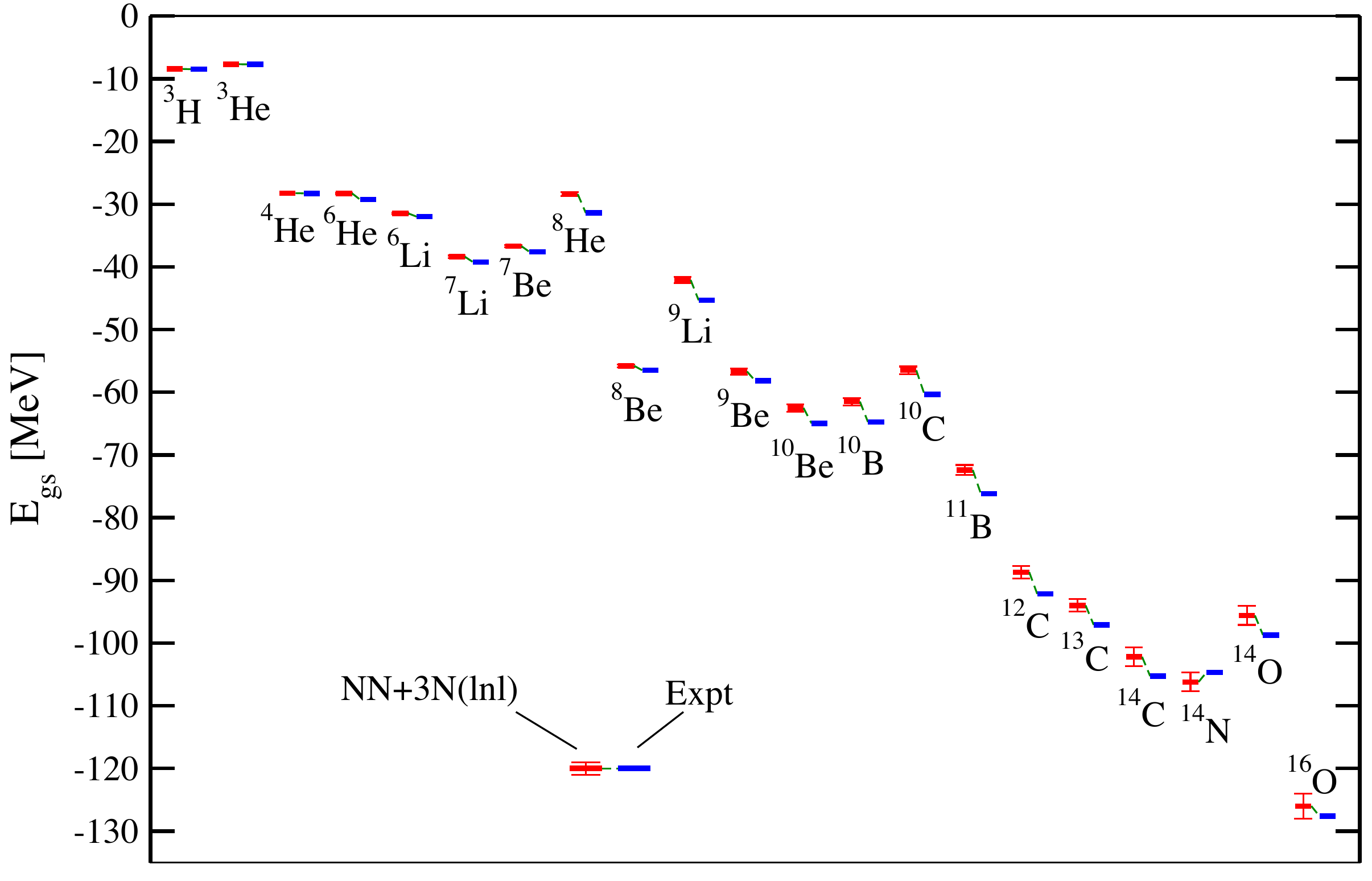}
\caption{Ground-state energies of $s$-shell and selected $p$-shell nuclei calculated with the \lnl{} Hamiltonian (red lines) compared to experiment (blue lines). 
The error bars indicate uncertainties of the NCSM extrapolation.
SRG evolution with $\lambda{=}2$ fm$^{-1}$ and HO frequency of $\hbar\Omega{=}20$ MeV were used.}
\label{fig_light_nuc_gs}
\end{figure}

As an initial test of the \lnl{} Hamiltonian, NCSM~\cite{Barrett2013} calculations of energies of $s$-shell and selected $p$-shell nuclei were performed. 
In the NCSM, nuclei are considered to be systems of $A$ nonrelativistic point-like nucleons interacting via realistic two- and three-body interactions. 
Each nucleon is an active degree of freedom and the translational invariance of observables, the angular momentum, and the parity of the nucleus are conserved. 
The many-body wave function is expanded over a basis of antisymmetric $A$-nucleon harmonic oscillator (HO) states. 
The basis contains up to $N_{\rm max}$ HO excitations above the lowest possible Pauli configuration, 
so that the motion of the center of mass is fully decoupled and its kinetic energy can be subtracted exactly. 
The basis is characterised by an additional parameter $\Omega$, the frequency of the HO well, and may depend on either Jacobi relative~\cite{navratil00fewN} or single-particle coordinates~\cite{Navratil2000}. 
The convergence of the HO expansion can be greatly accelerated by applying an SRG transformation on the $2N$ and $3N$ interactions~\cite{Wegner1994,Bogner2007,Hergert2007,Bogner2010,Jurgenson2009}.
Except for $A{=}3,4$ nuclei, here and in the following of the paper an SRG evolution is applied to the \ema{} and \lnl{} interactions down to a scale of $\lambda{=}2$~fm$^{-1}$.
On the contrary, calculations with \sat{} are performed with the bare Hamiltonian.

In Figs.~\ref{fig_Li}, \ref{fig_BeB} and \ref{fig_BC} the excitation energy spectra of selected Li, Be, B, and C isotopes are displayed.
A correct ordering of low-lying levels is found for all the considered lithium and beryllium isotopes, namely $^{6,7,9}$Li and $^{8,9}$Be. 
The $2^+ 0$ and $1^+_2 0$ states in $^6$Li as well as some of the \hbox{excited} states in $^7$Li and $^{8,9}$Be are broad resonances.
Here a more realistic description of $^6$Li and $^9$Be would require a better treatment of continuum effects, see Refs.~\cite{Hupin2015} and \cite{Langhammer2015}, respectively, in this regard. 
Let us note that all excited states of $^6$Li are unbound with respect to the emission of an $\alpha$ particle and that $^7$Li has only one excited state below the $\alpha$-separation threshold. 
Similarly, $^8$Be is never bound and even its ground state is unstable against decay into two $\alpha$.
The lowest states in $^{10}$B are known to be very sensitive to the details of nuclear forces, and the $3N$ interaction in particular~\cite{Navratil2007}. 
Here a good description is achieved by \lnl, with only the $1^+_2 0$ state resulting incorrectly placed.
The correct level ordering is also found in $^{11}$B, with the spectrum being overall too compressed as compared to the experimental one.
Finally, worth-noting is the correct ordering of $T{=}1$ states in $^{12}$C, also known to be sensitive to the $3N$ interaction. On the other hand, the alpha-cluster dominated $0^+ 0$ Hoyle state in $^{12}$C cannot be reproduced in the limited NCSM basis employed here~\cite{Epelbaum2011Hoyle}.
In general, \lnl{} yields spectra that are in good agreement with experiment. 
Some underestimation of level-splitting in $^{9}$Li, $^{11}$B, and $^{13}$C emerges, and could be associated with a weaker spin-orbit interaction strength. 
This is comparable to what has been found with earlier parameterisations of chiral $3N$ forces (see, e.g., Ref.~\cite{Navratil2007}). 

Ground-state energies of $^3$H, $^{3,4}$He, and selected $p$-shell nuclei from $^6$He to $^{16}$O are shown in Fig.~\ref{fig_light_nuc_gs}.
The calculated values (red lines) obtained with the \lnl{} interaction are compared to experiment (blue lines). 
Theoretical error bars represent the NCSM extrapolation uncertainty. 
Overall, experimental data are very reasonably reproduced, with differences of at most a few percent. 
The agreement is the best for $T_z{=}0$ and $|T_z|{=}1/2$ nuclei. 
Some deficiency of the interaction is observed with increasing $|T_z|$; for example, while $^4$He is in a perfect agreement with experiment, the $^6$He and $^8$He are barely bound.
Note however that, again, a proper treatment of continuum effects, not included here, is likely to provide additional binding to systems close to the dripline~\cite{Calci2016}.
Overall, the performance of the \lnl{} Hamiltonian in light nuclei is very encouraging.

\section{Medium-mass nuclei}
\label{sec_medium}

\subsection{Self-consistent Green's function theory}
\label{sec_GF}

In standard, i.e. Dyson, self-consistent Green's function theory (DSCGF)~\cite{Dickhoff04, Barbieri17}, the solution of the \hbox{$A$-body} Schr\"odinger equation is achieved via its rewriting in terms of one-, two-, ..., $A$-body objects named propagators or, indeed, Green's functions (GFs). Green's functions are expanded in a perturbative series, which in \textit{self-consistent} schemes is recast in terms of the exact GFs so that a large portion of non-skeleton diagrams are implicitly resummed. One is mostly interested in the one-body Green's function since this provides access to all \hbox{one-body} observables and to the ground-state energy via the so-called Galitskii-Migdal-Koltun sum rule~\cite{Galitskii:1958, koltun72}. The latter can be properly generalised to account for three-body forces~\cite{Carbone13}. In addition, the one-body GF contains information on neighbouring nuclei. Specifically, the residues from its Lehmann representation are related to transition matrix elements for one-nucleon addition and removal, while the poles give direct access to ground and excited states of ($A\pm1$)-nucleon systems. 
Note that in all calculations the intrinsic form of the Hamiltonian is employed, i.e. the center-of-mass kinetic energy is subtracted from the start. 
Since the latter depends on the number of nucleons at play, different calculations are performed with the Hamiltonian corresponding to mass number $A$ or $A\pm1$ depending on whether ground-state quantities or nucleon addition or removal spectra are computed, as detailed in Ref.~\cite{Cipollone15}.

The one-body GF is obtained by solving the Dyson equation that is intrinsically non-perturbative and in which the irreducible self-energy encodes all non-trivial many-body correlations arising from the interactions of a nucleon with the nuclear medium.  The self-energy is particularly important since it encodes information on both the $A$-nucleon ground state and the scattering states of the $A+1$ system. Hence it provides a natural \textit{ab initio} approach for consistent calculations of structure and reactions~\cite{Barbieri2005,Idini2019OpPot}.
In this work the self-energy is computed in the so-called algebraic diagrammatic construction [ADC($n$)] approach up to order $n$=3~\cite{Schirmer83, Barbieri17, Raimondi18}. This entails including all perturbative contribution up to $n$-th order plus any additional resummation needed to preserve its spectral representation. At first order, ADC(1) includes only mean-field terms and it is nothing else than the standard Hartree-Fock (HF) approximation. 
The higher orders, ADC(2) and ADC(3), add dynamical correlations in terms of 2p1h and 2h1p configurations. However, these remain minimally included and non-interacting at ADC(2) while ADC(3) includes infinite-order resummations of both particle-particle/hole-hole and particle-hole ladders.  Generally speaking, ADC($n$) defines a truncation scheme that is systematically improvable up to ADC($\infty$), where exact results are recovered by definition.

In  Dyson GF theory the diagrammatic expansion builds on top of a reference state that is particle-number conserving and that typically respects spherical symmetry. While such an expansion can suitably address doubly closed-shell systems, it becomes inefficient or even breaks down in open-shell systems due to the degeneracy of the reference state with respect to particle-hole excitations. With the wish to retain the simplicity of a single-reference method, a possible solution consists in working, from the outset, with a symmetry-breaking reference state.  In particular, breaking U(1) symmetry associated with particle number conservation\footnote{In the case of atomic nuclei proton and neutron numbers are conserved individually, and therefore it is always intended U(1)$_N \otimes $U(1)$_Z$ where one of the two or both are broken.} while maintaining spherical symmetry allows us to efficiently capture pairing correlations, thus gaining access to (singly) open-shell nuclei. 

In this spirit, Ref.~\cite{gorkov}  generalised DSCGF to a U(1) symmetry-breaking scheme based on the use of a Hartree-Fock-Bogolyubov reference state and we refer to this approach as Gorkov self-consistent Green's function (GSCGF) theory. The resulting four (two normal and two anomalous) Gorkov propagators can be conveniently recast in a $2 \times 2$ matrix notation via Nambu formalism~\cite{Nambu60}. Hence, all standard GF equations are rewritten in a Nambu-Gorkov matrix form. Moreover, Dyson diagrammatics can be generalised to a Gorkov framework, with minor complications arising from the presence of the four different one-body propagators~\cite{Soma11a}.  The introduction of a chemical potential guarantees that the number of particles is the correct one \textit{on average}. Eventually, the broken symmetry has to be restored. While symmetry-restored formalism has been developed for other (post-Hartree-Fock) many-body methods~\cite{Duguet:2014jja, Duguet17b, Qiu2017PrCC1,Qiu2018PrCC2,Qiu:2018edx}, it remains to be formulated for GSCGF.
GSCGF theory has been recently implemented in the context of nuclear physics within the ADC(2) truncation scheme~\cite{Soma11a, Soma13, Soma14a} . Thus, the present paper reports results of DSCGF calculations up to ADC(3) and GSCGF calculations up to ADC(2)\footnote{When closed-shell systems are considered, a Gorkov calculation automatically reduces to a Dyson one.}. 

\subsection{Model-space convergence}
\label{sec_MSconv}

\begin{figure}[t]
\centering
\includegraphics[width=9.2cm]{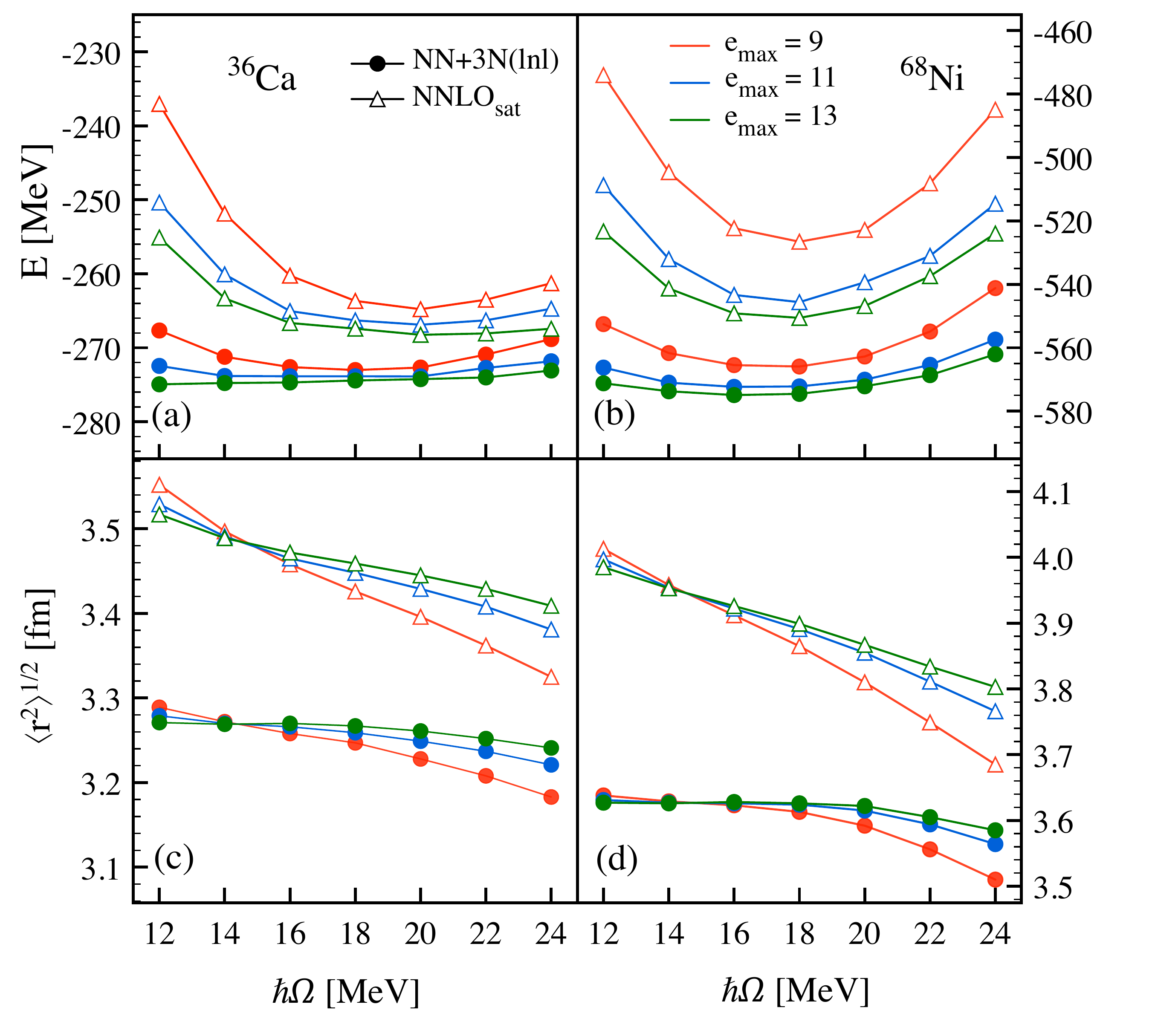}
\caption{Ground-state energy computed at the ADC(2) level with the \lnl{} and \sat{} Hamiltonians as a function of the harmonic oscillator spacing $\hbar\Omega$ and for increasing size $e_{\text{{\rm max}}}$ of the single-particle model space for the cases of $^{36}$Ca, panel (a), and $^{68}$Ni, panel (b).
Rms radii from the same calculations are displayed for $^{36}$Ca and $^{68}$Ni in panels (c) and (d) respectively.}
\label{fig_NmaxHW}
\end{figure}

The present calculations  are performed using a spherical HO model space that includes up to the $e_{\text{{\rm max}}} \equiv$~max~$(2n+l)$ = 13 shell. Any $k$-body operator is to be represented in the same space and one should truncate the corresponding $k$-body basis consistently according to $e_{k\text{max}} = k\,e_{\text{max}}$.  
The matrix elements of one- and two-body operators are always included in full. 
However, this is not feasible for three-body interactions due to the rapid increase in the number of their matrix elements and therefore these are restricted to $e_{3\text{max}}=16<3\,e_{\text{max}}$.
The dependence on the basis parameters was tested by computing ground-state observables for different harmonic oscillator frequencies, $\hbar \Omega$, and model space sizes, $e_{\text{max}}$.
Results for ground-state energies, $E$, and root-mean-square~(rms) radii, $\langle r^2_{\text{ch}} \rangle^{1/2}$, computed at the ADC(2) truncation level are displayed in Fig.~\ref{fig_NmaxHW} for two representative nuclei, $^{36}$Ca and $^{68}$Ni, and for \lnl{} and \sat{} Hamiltonians.

Focusing on ground-state energies, both interactions show a typical convergence pattern consisting in curves that gradually become independent of $\hbar \Omega$ and closer to each other as the basis increases. 
For both nuclei, the change from $e_{\text{max}}=9$ to $e_{\text{max}}=13$ at the $\hbar \Omega$ minimum is larger for \sat{} than for \lnl{}, consistently with the SRG-evolved character of the latter.
In $^{36}$Ca, going from $e_{\text{max}}=11$ to $e_{\text{max}}=13$ results in a 1.5 MeV gain for \sat{} and a 300 keV gain for \lnl. In $^{68}$Ni gains are 4.9 and 2 MeV respectively.
Note that the basis limitations on three-body forces do not affect the lighter systems considered in this work and $e_{3\text{max}}$ = 16 is normally sufficient to converge isotopes around $^{40,48}$Ca. However, this truncation can introduce some uncertainties for heavier masses. We shall quantify these errors in Sec.~\ref{sec_GS}, when discussing the neutron-rich nuclei in the $pf$-shell.
Charge radii also show their usual convergence pattern, with the $\hbar \Omega$ dependence decreasing as the model space truncation increases.
For $e_{\text{max}}=13$ calculations, the \lnl{} curves are already rather flat in an interval including both the energy minimum and smaller values of $\hbar \Omega$.
Results with \sat, on the other hand, still present a manifest $\hbar \Omega$ dependence. 
Hence, a precise determination of the rms charge radius would require the use of extrapolation techniques. 
As the conclusions of the large-scale systematic analyses presented in this work would not be impacted, such extrapolation is not performed here and is left for future studies.

\subsection{Many-body convergence}
\label{sec_MBconv}

\begin{figure}
\centering
\includegraphics[width=9cm]{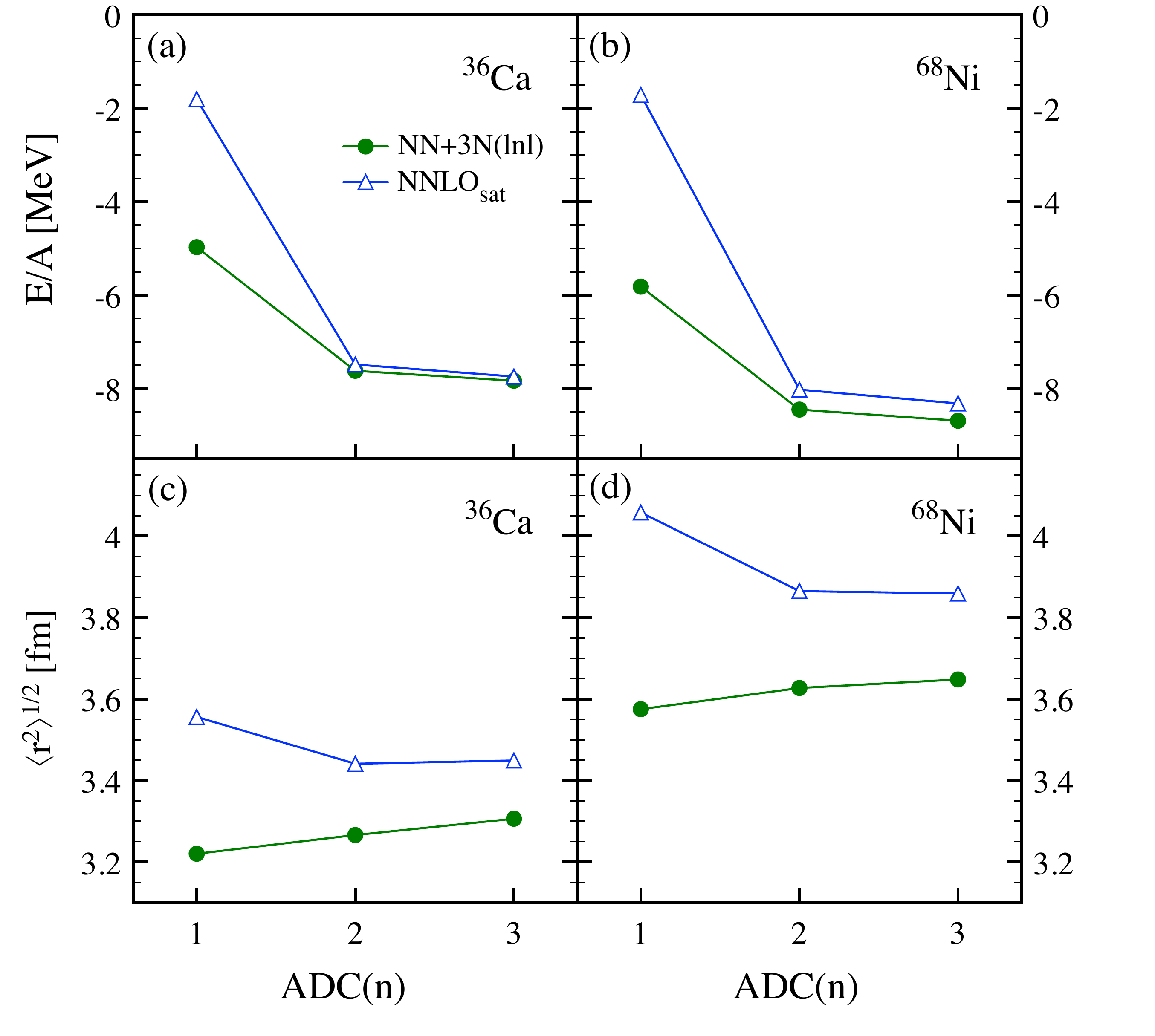}
\caption{Ground-state energies per particle (top panels) and rms charge radii (bottom panels) of $^{36}$Ca and $^{68}$Ni computed within different ADC($n$) truncation schemes.
Results for the \sat{} and \lnl{} interactions are displayed.}
\label{fig_ADC_convergence}
\end{figure}

Next, let us investigate convergence with respect to the many-body truncation. 
In Fig.~\ref{fig_ADC_convergence}, energies per nucleon and rms charge radii computed within ADC(1), ADC(2) and ADC(3) approximations are displayed for the same two representative cases of $^{36}$Ca and $^{68}$Ni.
A clear convergence pattern is visible in all cases.
For ground-state energies, ADC(1) results depend strongly on the interaction, with the softer \lnl{} more bound than \sat.
ADC(2) calculations are already sufficient to grasp the bulk of correlation energy.
Interestingly, ADC(2) values from the two interactions are similar, which reflects the ability of this self-consistent scheme to resum relevant many-body contributions even in the presence of an SRG-unevolved (though relatively soft) interaction.
Going from ADC(2) to ADC(3) results in a further gain in correlation energy, which shows that the ADC(3) truncation level is necessary for precise estimates of total ground-state energies even when evolved or soft interactions are employed.
Quantitatively, when going from ADC(2) to ADC(3), one gains 7.6 MeV (9.2 MeV) absolute energy in $^{36}$Ca and 16.2 MeV (20.1 MeV) in $^{68}$Ni with \lnl{} (\sat)\footnote{This corresponds to a 7.9$\%$ (4.5$\%$) increase of correlation energy in $^{36}$Ca and a 9.0$\%$ (4.7$\%$) increase in $^{68}$Ni with \lnl{} (\sat).}. Extrapolating this convergence sequence, one may expect ADC($n$) with $n \geq 4$ to add as little as $\sim 1\%$ correlation energy.
As discussed further in Secs.~\ref{sec_GS} and ~\ref{sec_radii}, while corrections from ADC(3) are important when confronting total ground-state energies they tend to remain rather constant across whole isotopic chains. This implies that ADC(2) already yields reliable predictions for trends and differential quantities such as two-nucleon separation energies.  On the other hand, the ADC(3) is also known to be important to reproduce affinities and ionization spectra in molecules~\cite{vonNiessen1984QCadc} and, correspondingly, it gives significant corrections to the spectra of dominant quasiparticle states discussed in Sec.~\ref{sec_spectra}.

In the case of calculations performed in the Gorkov framework, an additional source of error comes from the fact that the broken U(1) symmetry is not presently restored.
While the number of particles remains the correct one on average, this leads to a dispersion in $N$ and/or $Z$ depending on the open-shell nature of neutron or protons.
Here only calculations in semi-magic nuclei are reported, for which the proton variance remains zero.
At the ADC(2) level, the maximum variance in neutron number amounts to $\sigma_N^2 \approx$ 1.2, 1.6 and 1.9 for oxygen, calcium and nickel chains respectively (independently of the employed interaction).

For rms radii the convergence pattern results are even more favourable. While ADC(1) already provides a reasonable account of the charge radius,  ADC(2) is necessary to reach an essentially converged value, especially in the case of \sat. Eventually, ADC(3) adds at most 1.2$\%$ to the rms radii that is well converged with respect to the many-body truncation for both \lnl{} and \sat.
The above findings on both spectra and radii are consistent with the analysis performed on $^{34}$Si and $^{36}$S in Ref.~\cite{Duguet17a}. The same formalism applied in the context of quantum chemistry also showed a similar ADC convergence behaviour~\cite{vonNiessen1984QCadc,Danovich2011,Degroote2011frpa,Barbieri2012PRA}.

 \begin{table}[t]
\begin{center}
\begin{ruledtabular}
\begin{tabular}{lccc}
  $E_{g.s.}$ ($^{16}$O):        &  ADC(2)  & ADC(3)  &  NCSM    \\
\hline
\ema      & -128.54  &  -130.81  & -130(2)  \\
\sat        & -124.63  &  -126.23  & -125(5)  \\
\lnl         & -123.91  &  -127.27  & -126(2)  \\
\hline
\multicolumn{3}{l}{Experiment:} & -127.62
\end{tabular}
\end{ruledtabular}
\caption{Comparison between the ground state energies of $^{16}$O (in MeV) as computed with the ADC(2) and ADC(3) many-body approximations and with the NCSM. 
The SCGF results are obtained in a full $e_{\text{max}}=13$ space and one could expect residual errors of a few \% due to the many-body truncations beyond ADC(3), as detailed in the text. 
The NCSM results for \sat{} and \lnl{} show uncertainties arising from $N_{\rm max}$ extrapolation and importance-truncation.
The NCSM result for \ema{} is taken from Ref.~\cite{Hergert13}.
\label{O16bench} }
\end{center} 
\end{table}
 Table~\ref{O16bench} displays a benchmark between the NCSM and SCGF for $^{16}$O and the three Hamiltonians.
This isotope is still light enough that it can be computed using the importance-truncated NCSM~\cite{Roth2007,Roth2009}, although the presence of matrix elements from the $3N$ interaction limits the largest possible basis truncation to \hbox{$N_{\rm max}=10$.}~
 This issue is more severe for the harder \sat{} interaction. 
 Thus, the NCSM results have been extrapolated using a standard exponential trend with respect to $N_{\rm max}$ and a polynomial dependence on the importance-truncation parameter $\kappa_{min}$~\cite{Roth2007}.
Table~\ref{O16bench} reports the extrapolated values together with the uncertainties estimated from this procedure by repeating the extrapolations with different subsets of data.
Taking these uncertainties under consideration,  the comparison among the two \emph{ab initio} approaches is extremely satisfactory and confirms the reliability of SCGF for the computation of medium-mass isotopes presented in the following.  From Table~\ref{O16bench}, one also notices that the \ema---which was the first Hamiltonian to successfully predict the oxygen ground state energies \textit{ab initio}---already displays a slight tendency to overbind, even for this nucleus. The \sat{} and \lnl{} correct this effect, with the latter performing a bit better in comparison to the experiment.

\subsection{Ground-state energies}
\label{sec_GS}

The following subsections study the performances of the three Hamiltonians --\ema, \lnl{} and \sat-- along three representative medium-mass isotopic chains, namely oxygen, calcium and nickel.
Based on the considerations of Secs.~\ref{sec_MSconv} and~\ref{sec_MBconv}, all  the following calculations are performed with an $e_{\text{max}}=13$ model space (14 shells), $e_{3\text{max}}$ = 16, and oscillator frequencies fixed at $\hbar \Omega=20$ MeV for \sat{} and $\hbar \Omega=18$ MeV for \lnl{}.
Similar studies have shown that \ema{} has an optimal minimum at $\hbar \Omega=28$ MeV~\cite{Soma14a}, which is used here for this Hamiltonian.
For the three isotopic chains, protons maintain a good closed-shell character, i.e. all isotopes are at least semi-magic, which generally ensures that deformation does not play a major role\footnote{A possible exception is represented by some nickel isotopes between $^{56}$Ni and $^{68}$Ni, as discussed later.}. 
Ground-state properties (total binding energies, charge radii, density distributions) of even-even nuclei as well as excitation spectra of odd-even nuclei are investigated to provide a comprehensive  benchmark of the three interactions.

\begin{figure}[t]
\centering
\includegraphics[width=8.6cm]{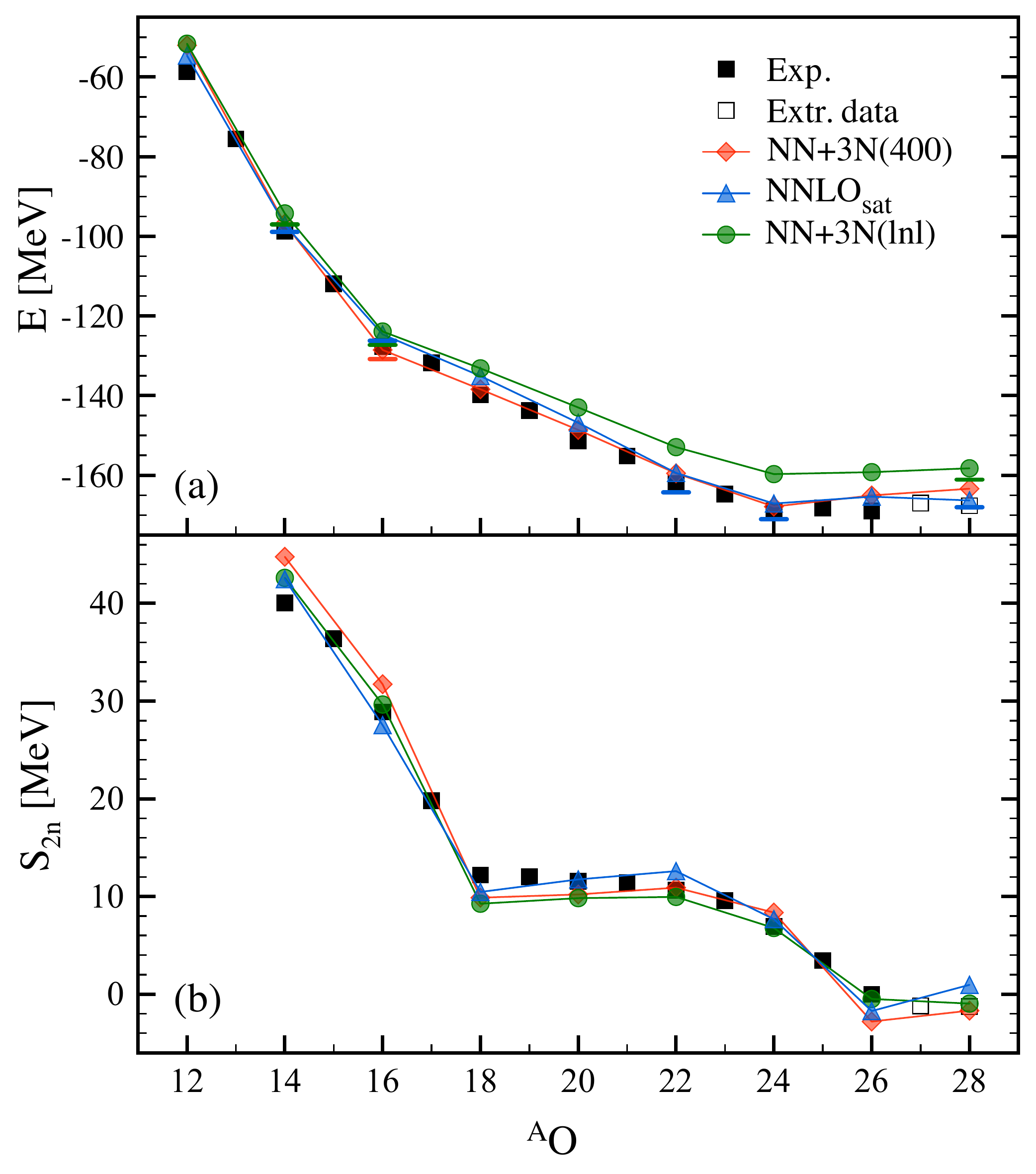}
\caption{Total binding energies (a) and two-neutron separation energies (b) of oxygen isotopes computed within the ADC(2) approximation with the \ema, \lnl{} and \sat{} interactions. 
ADC(3) calculations with the \lnl{} and \sat{} interactions are also displayed for closed-shell nuclei as horizontal bars. 
Calculations are compared to measured as well as extrapolated data from the 2016 atomic mass evaluation (AME)~\cite{AME2016, Wang17}.
The estimated computational errors due to model space truncations are below 1\% of the total binding energy for \sat{} and below 0.5\% for \lnl{} and 
\ema{}. Note that the ADC(3) truncation accounts for an additional 2-3\% of the total binding energies with respect to ADC(2), for all interactions and throughout this chain.
}
\label{fig_BE_O}
\end{figure}
\begin{figure}
\centering
\includegraphics[width=8.6cm]{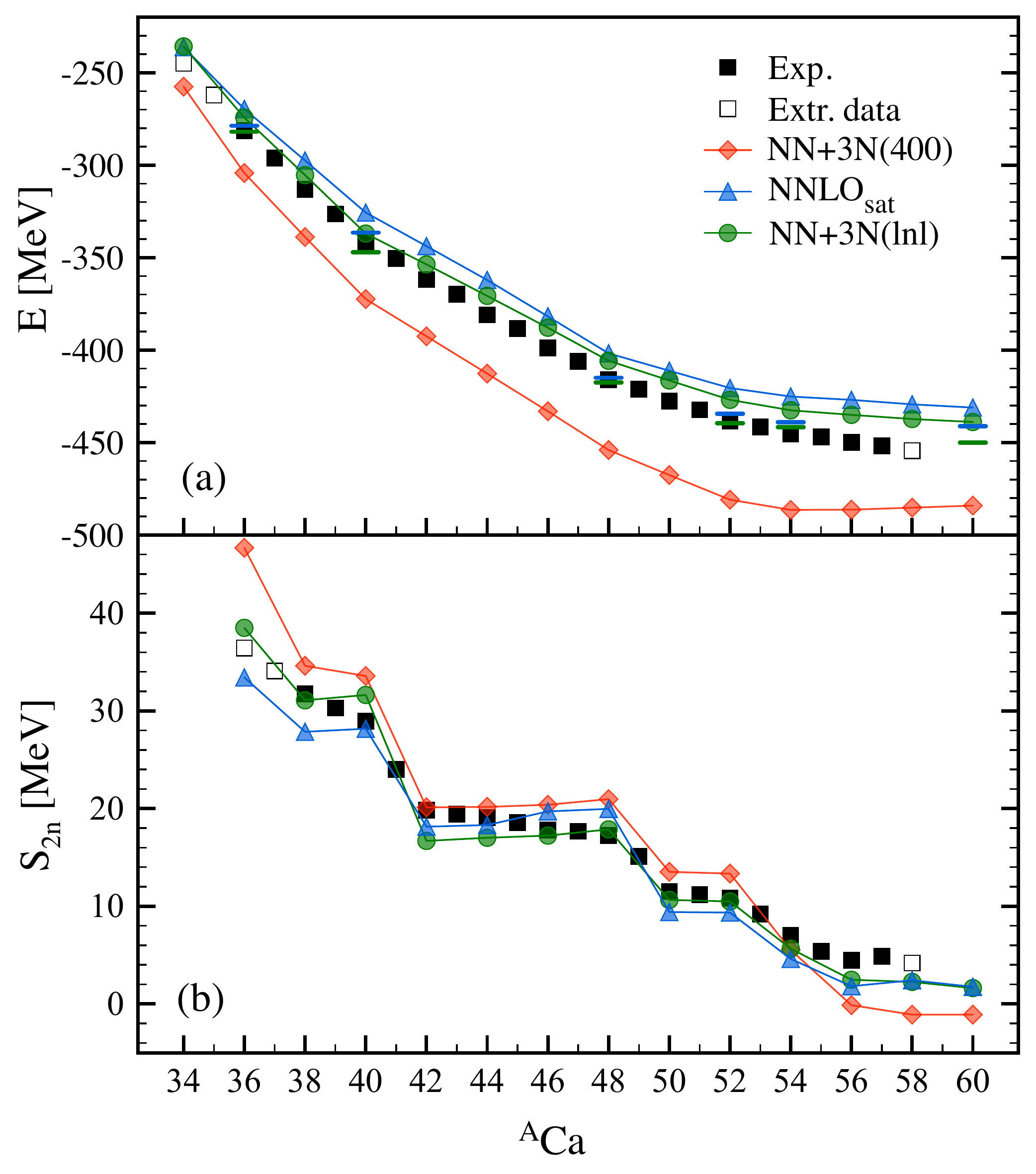}
\caption{Same as Fig.~\ref{fig_BE_O} but for calcium isotopes. Values for the recently measured masses of $^{55-57}$Ca were taken from Ref.~\cite{Michimasa18}.
The estimated computational errors due to model space truncations are $\approx$ 1\% of the total binding energy for \sat{} and 0.5\% for \lnl{} and \ema{}.}
\label{fig_BE_Ca}
\end{figure}

\begin{figure}
\centering
\includegraphics[width=8.6cm]{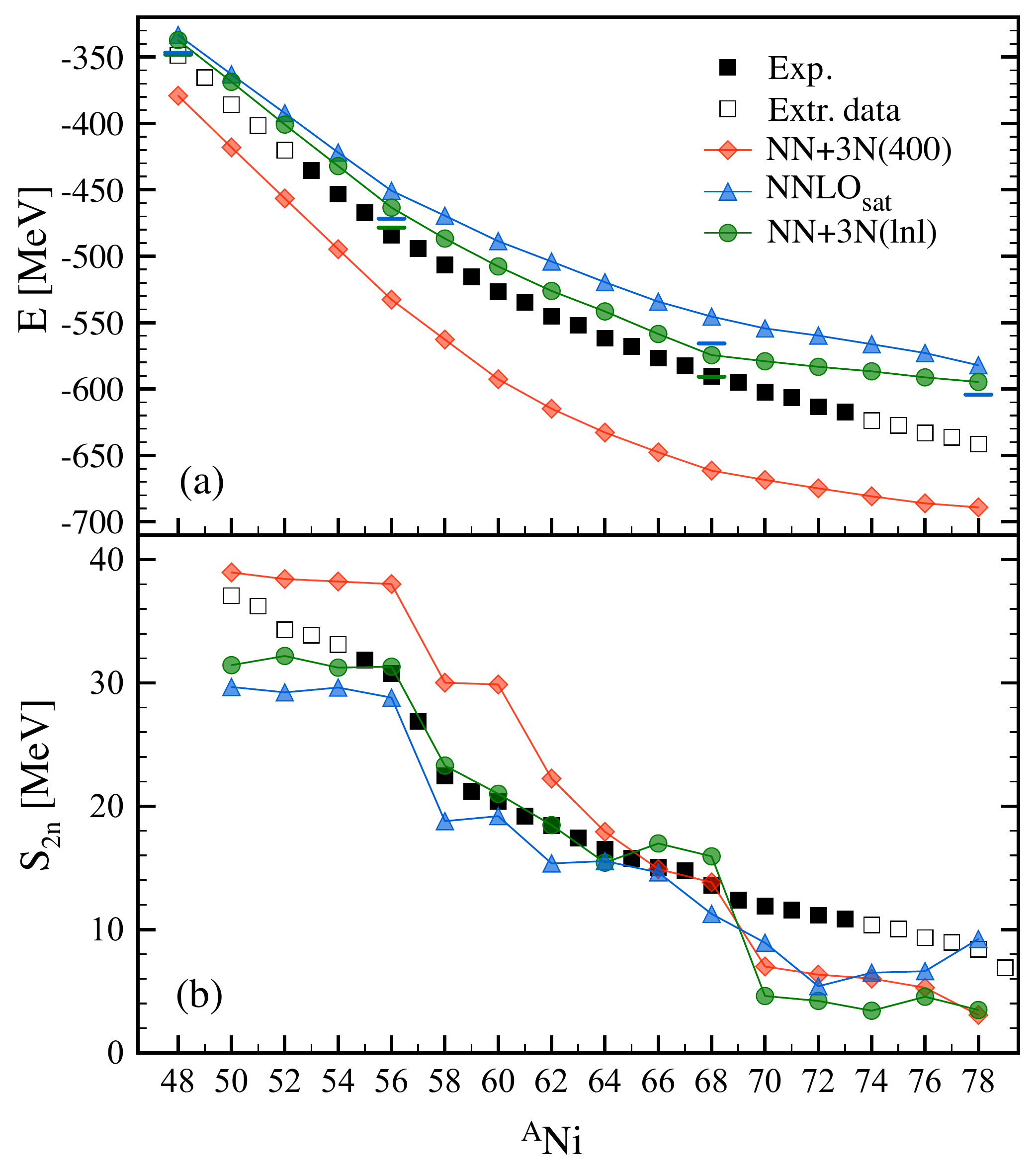}
\caption{Same as Fig.~\ref{fig_BE_O} but for nickel isotopes.
The estimated computational errors due to model space truncations are below 2\% of the total binding energy for \sat{} and below 1\% for \lnl{} and 
\ema{}. Note that the ADC(3) truncation accounts for an additional 2-3\% of the total binding energies with respect to ADC(2), for all interactions throughout this chain.}
\label{fig_BE_Ni}
\end{figure}

Total ground-state energies of oxygen, calcium and nickel isotopes are displayed in Figs.~\ref{fig_BE_O}(a), \ref{fig_BE_Ca}(a) and \ref{fig_BE_Ni}(a) respectively.
ADC(2) results (coloured points and lines), covering all even-even isotopes, are shown together with ADC(3) calculations in doubly closed-shell nuclei (coloured horizontal bars) and compared to available experimental data (black points).
Corresponding two-neutron separation energies are shown in panels~(b).
Following the analysis of Secs.~\ref{sec_MSconv} and~\ref{sec_MBconv}, model-space convergence errors for \sat{}~(\lnl{}) are estimated to be at most 1\%~(0.5\%) of the total binding energy up to the calcium isotopes and 2\%~(1\%) for the nickels up to $^{68}$Ni. 
Many-body truncation errors are 4\% for ADC(2) and below 1\% for ADC(3), generally underestimating the binding energy. 
Uncertainties for \ema{} are the same as for \lnl{}.

All three interactions yield similar results for ground-state energies of the oxygen isotopes and are generally close to experimental values.
While for \ema{} and \sat{} the agreement is excellent through the whole chain, \lnl{} shows some mild underbinding for the most neutron-rich systems.
Although additional correlations coming in at the ADC(3) level tend to provide additional binding, one notices that this effect is not large in oxygen.
For all interactions the dripline at $^{24}$O is correctly reproduced, as also visible in Fig.~\ref{fig_BE_O}(b).
For the model space parameters used here, the two N$^3$LO Hamiltonians predict $^{28}$O to be less bound than $^{26}$O, while the opposite is found for \sat{}. However, we find that computed ground-state energies for the unbound $^{28}$O depend sensibly on $e_{\text{max}}$ and $\hbar \Omega$ which is consistent with a discretization of the continuum imposed by the HO space.

For heavier systems like calcium and nickel, the \ema{} Hamiltonian is known to produce strong overbinding with respect to experimental data~\cite{Binder14, Soma14b}.
This is confirmed by present calculations as visible in Figs.~\ref{fig_BE_Ca}(a) and \ref{fig_BE_Ni}(a).
Instead, one notices that the two most recent Hamiltonians, \sat{} and \lnl, largely correct for this overbinding.
For instance, on the light-mass side, the ADC(2) energy for $^{36}$Ca goes from 20.4 MeV (7.2$\%$) overbinding for \ema{} to 11.8 MeV (4.1$\%$) underbinding for \sat{} and 7.0 MeV (2.4$\%$) underbinding for \lnl.
Among the heavier isotopes,  $^{68}$Ni  goes from 64.8 MeV (10.9$\%$) overbinding for \ema{} to 45.0 MeV (7.6$\%$) underbinding for \sat{} and 15.9 MeV (2.6$\%$) underbinding for \lnl.

Many-body correlations beyond ADC(2) provide additional binding
and ground-state energies of all considered isotopes are lower by 2-3\% when switching to ADC(3). 
While this aggravates the overbinding of \ema~\cite{Soma14b}, it is expected to reduce the underbinding of the other two potentials. 
The latter expectation is corroborated by ADC(3) results of closed-shell nuclei along the two chains. 
Once ADC(3) corrections are included, binding energies computed with both \sat{} and \lnl{} Hamiltonians are in excellent agreement with experimental data.  For the above examples, differences with experiment reduce to 0.9$\%$ and 0.2$\%$ in $^{36}$Ca and to 4.2$\%$ 
and 0.05$\%$ in $^{68}$Ni for \sat{} and \lnl{} respectively
and are comparable with the theoretical uncertainties due to the model-space convergence.
The extra binding obtained within ADC(3) is therefore crucial if one is after precise comparisons on total ground-state energies.
In accordance with the findings of Sec.~\ref{sec_MBconv}, 
ADC(3) corrections are systematically larger for the SRG-unevolved \sat{} than for the SRG-evolved \lnl{}.
Note that Ref.~\cite{Hagen2016prlNi78} already reported very poor convergence of the  $^{78}$Ni isotope with \sat{}, mainly due by the truncation of three-body matrix elements. 
Here, it is found that changing $e_{3\text{max}}$ from 14 to 16 leads to a $\approx$40~MeV variation, which adds to the model space uncertainties discussed in Sec.~\ref{sec_MSconv}. 
All isotopes beyond $^{68}$Ni are likely to be affected in an analogous way.

Further insight can be gained by looking at energy differences.
Two-neutron separation energies along the two chains are displayed in Figs.~\ref{fig_BE_Ca}(b) and \ref{fig_BE_Ni}(b).
For calcium, the effects of overbinding in \ema{} cancels out to a large extent, with the residual mass dependence showing up in the most proton- and neutron-rich isotopes.
Calculations with \sat{} and \lnl{} lead to similar values and closely follow experimental data.
Interestingly, all major gaps correctly emerge.
While the $N=20$ gap is quantitatively well reproduced by \sat, it appears somewhat overestimated with \lnl.
The opposite holds for the $N=28$ gap, with \lnl{} providing a very accurate description.
\begin{figure}
\centering
\includegraphics[width=9.1cm]{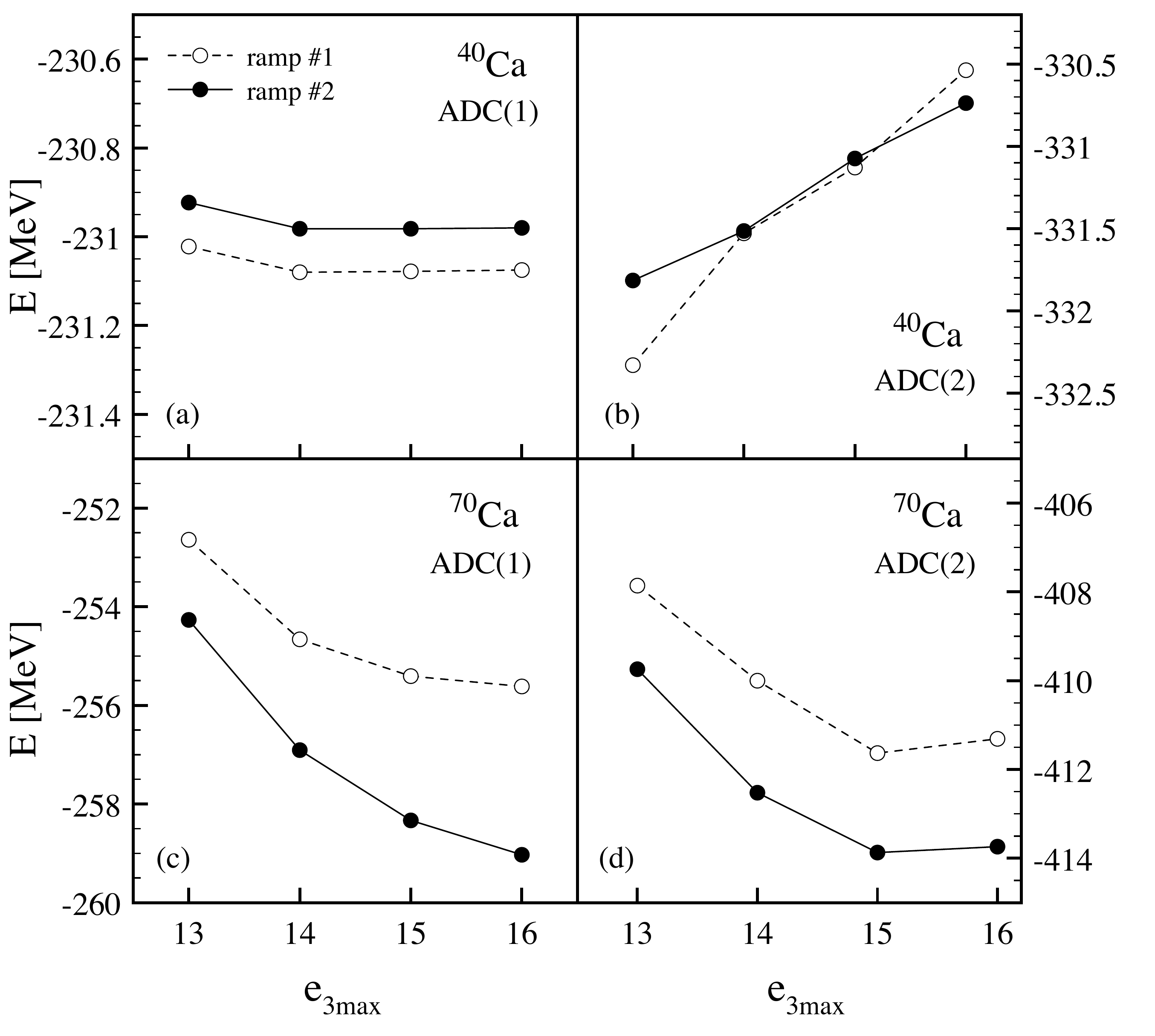}
\caption{Ground-state energies of $^{40}$Ca (upper panels) and $^{70}$Ca (lower panels) for \lnl{} as a function of the size of the three-body basis $e_{3\text{max}}$. Calculations are shown for two different truncation schemes used in SRG evolution: `ramp \# 1' refers to including Jacobi states with $N_{\text{max}}$$\geq$30 up to  $J$=25/2 and $N_{\text{max}}$=16 up to $J$=35/2, while `ramp \# 2' retains $N_{\text{max}}$=30 states up to  $J$=31/2. Note the different energy scales in the upper and lower panels.}
\label{fig_ramp}
\end{figure}
One compelling question, attracting much attention both experimentally and theoretically, relates to the position of the dripline in calcium isotopes.
Very recently, the experimental knowledge was extended with the mass measurements of $^{55-57}$Ca~\cite{Michimasa18} and the first evidence of a bound $^{60}$Ca~\cite{Tarasov18}.
While the present theoretical framework is in principle able to be applied beyond $^{60}$Ca, convergence problems were encountered with the present settings and the (inconclusive) results beyond $^{60}$Ca are not shown here. 
In particular, the SRG evolution of three-body operators is performed in a three-body HO space in Jacobi coordinates~\cite{navratil00fewN} and requires very high total angular momenta (up to $J$=35/2) to resolve all matrix elements up to $e_{3\text{max}}$ = 16 used in the SCGF model space. 
Calculations for isotopes above $^{60}$Ca were found to be affected by model space truncations in two different ways: one is the direct dependence on $e_{3\text{max}}$ in the SCGF basis
and the other is the truncations needed to evolve \lnl{} through SRG. 
As an example, in Fig.~\ref{fig_ramp} total energies of $^{40}$Ca and $^{70}$Ca are displayed as a function of $e_{3\text{max}}$ for two different ways of truncating the Jacobi basis during the SRG evolution.
While calculations in $^{40}$Ca are converged with respect to both variations, in $^{70}$Ca a clear dependence on the latter parameter is visible.
This problem arises only beyond $N=40$ for the two SRG-evolved Hamiltonians and is also responsible for a lack of binding in neutron-rich nickel isotopes, which is reflected in an evident kink in the binding energy curve after $^{68}$Ni (see Fig.~\ref{fig_BE_Ni}).
Future technical improvements as well as a better treatment of the continuum, which appears to be crucial beyond $^{60}$Ca, are therefore necessary for a correct determination of the calcium dripline.

Two-neutron separation energies in the nickel chain are generally less accurate.
The old \ema{} Hamiltonian struggles to catch the experimental trend, with an unrealistic large gap appearing at $N=32$. 
The other two interactions show a clear improvement.
Interestingly, both of them predict a flat trend for proton-rich isotopes, in contrast with the AME data extrapolation.
For \lnl, the agreement with experiment between $^{56}$Ni and $^{64}$Ni is remarkable. 
As explained above, after $^{68}$Ni the results appear to be affected by convergence issues.
For \sat, the description remains reasonable except for most neutron-rich systems.
The reproduction of experimental data also deteriorates between $N=28$ and $N=40$, which is likely to be linked with the onset of deformation although this does not appear to be problematic for the soft \lnl{} interaction.

\subsection{Charge radii and density distributions}
\label{sec_radii}

\begin{figure}[t]
\centering
\includegraphics[width=8.6cm]{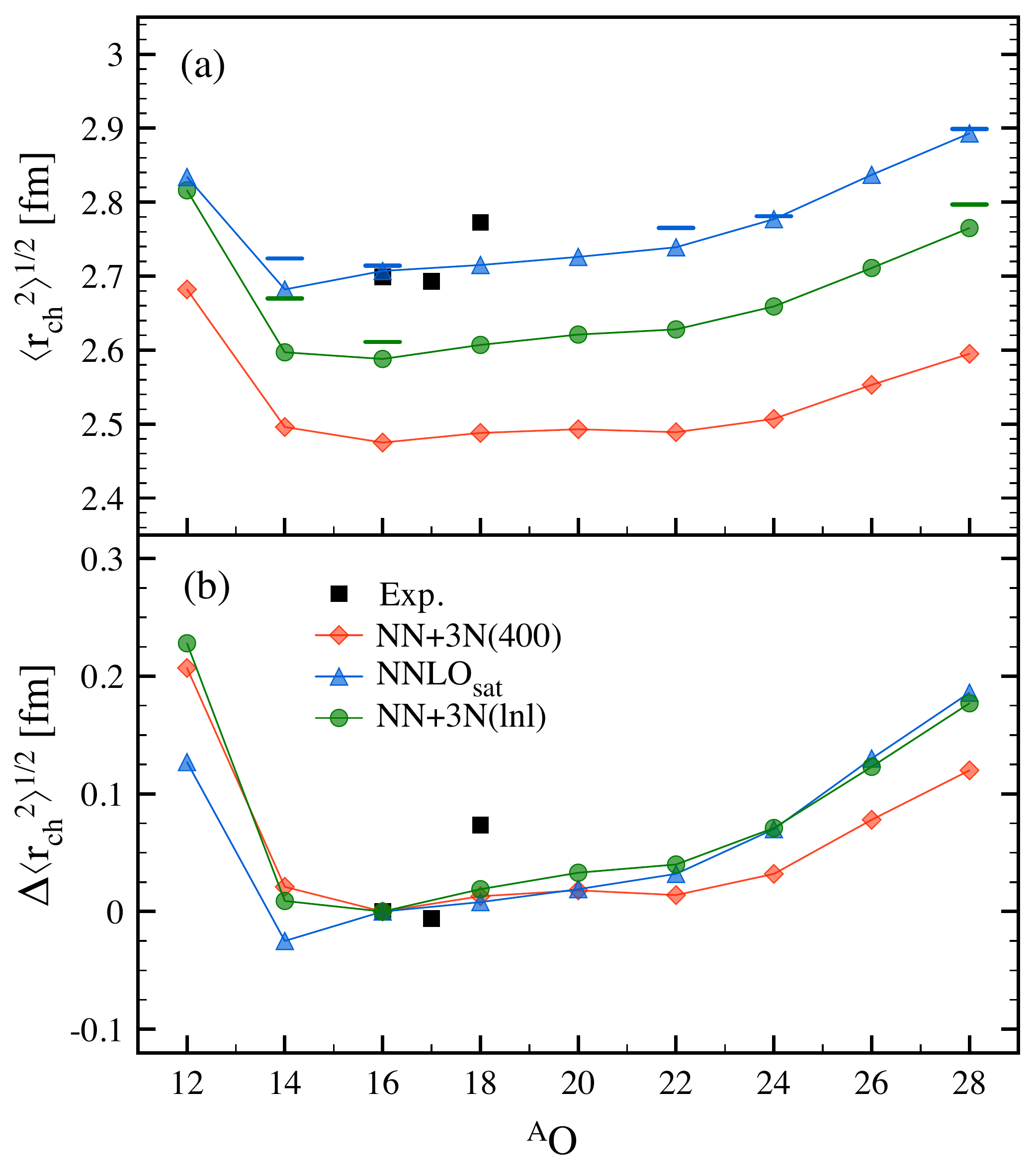}
\caption{Rms charge radii of oxygen isotopes (a) and differential radii relative to $^{16}$O (b) computed within the ADC(2) approximation with the \ema, \lnl{} and \sat{} interactions. ADC(3) calculations with the \lnl{} and \sat{} interactions are shown for closed-shell nuclei as horizontal bars. Available experimental data~\cite{Angeli13} are also displayed.
The estimated computational errors due to model space truncations are less than 1.5\% of the charge radius for \sat{}, while calculations are
substantially converged for  \lnl{} and  \ema{} (see Fig.~\ref{fig_NmaxHW}).
The corrections to radii due to many-body truncation beyond ADC(3) are larger for \lnl{} than for \sat{} and remain $<$0.01~fm in all cases (see Fig.~\ref{fig_ADC_convergence}).}
\label{fig_radii_O}
\end{figure}

\begin{figure}
\centering
\includegraphics[width=8.6cm]{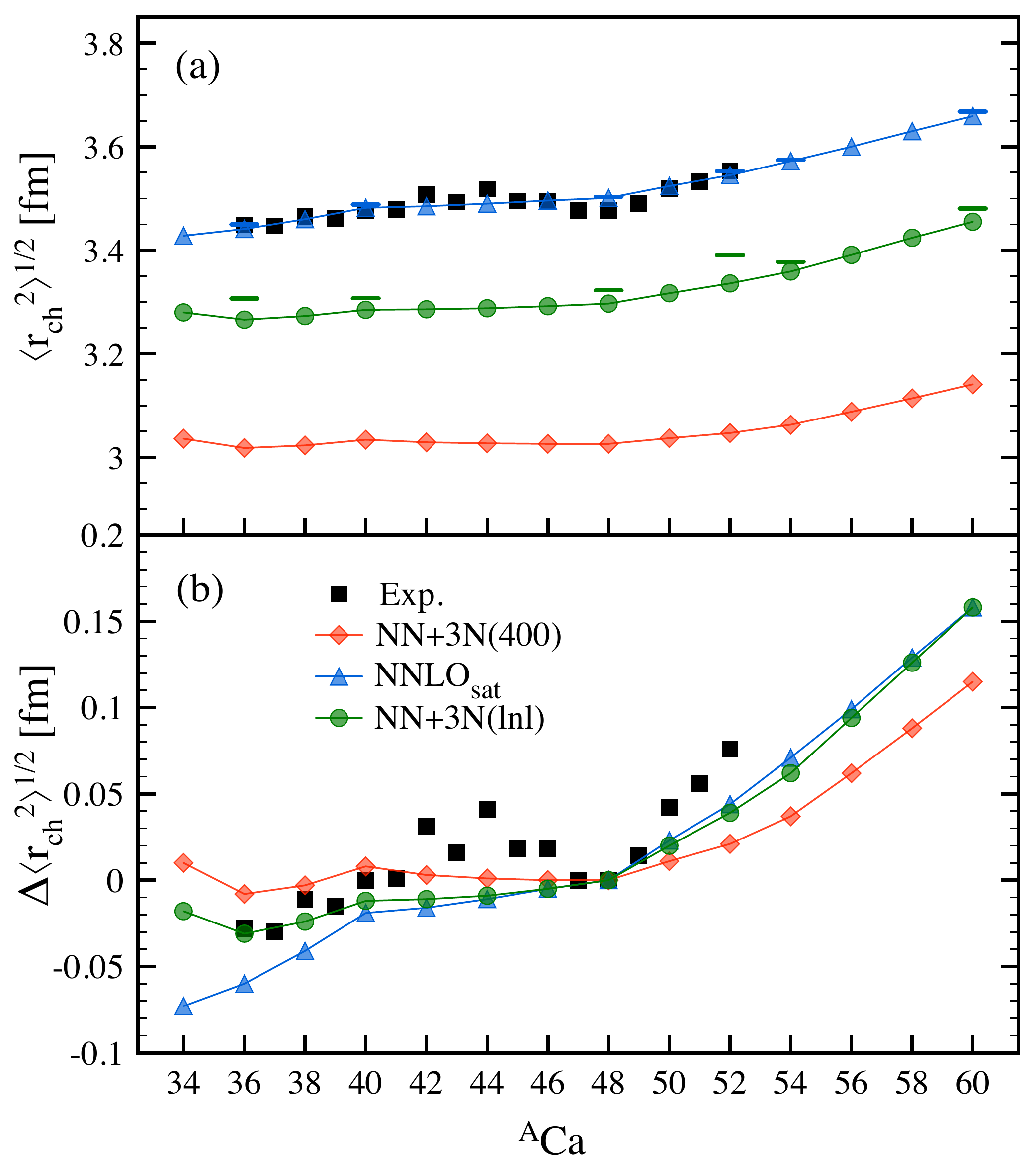}
\caption{Same as Fig.~\ref{fig_radii_O} but for calcium isotopes. In panel (b), differential radii are relative to $^{48}$Ca. Experimental data are taken from Refs.~\cite{Angeli13, Ruiz16, Miller19}.
The estimated computational errors due to model space truncations are $\approx$ 1.5\% of the charge radius for \sat{}, while calculations are
substantially converged for \lnl{} and  \ema{} (see Fig.~\ref{fig_NmaxHW}). The corrections to radii due to many-body truncation
beyond ADC(3) are larger for \lnl{} than for \sat{} and remain $<$0.01~fm in all cases (see Fig.~\ref{fig_ADC_convergence}).}
\label{fig_radii_Ca}
\end{figure}

\begin{figure}
\centering
\includegraphics[width=8.6cm]{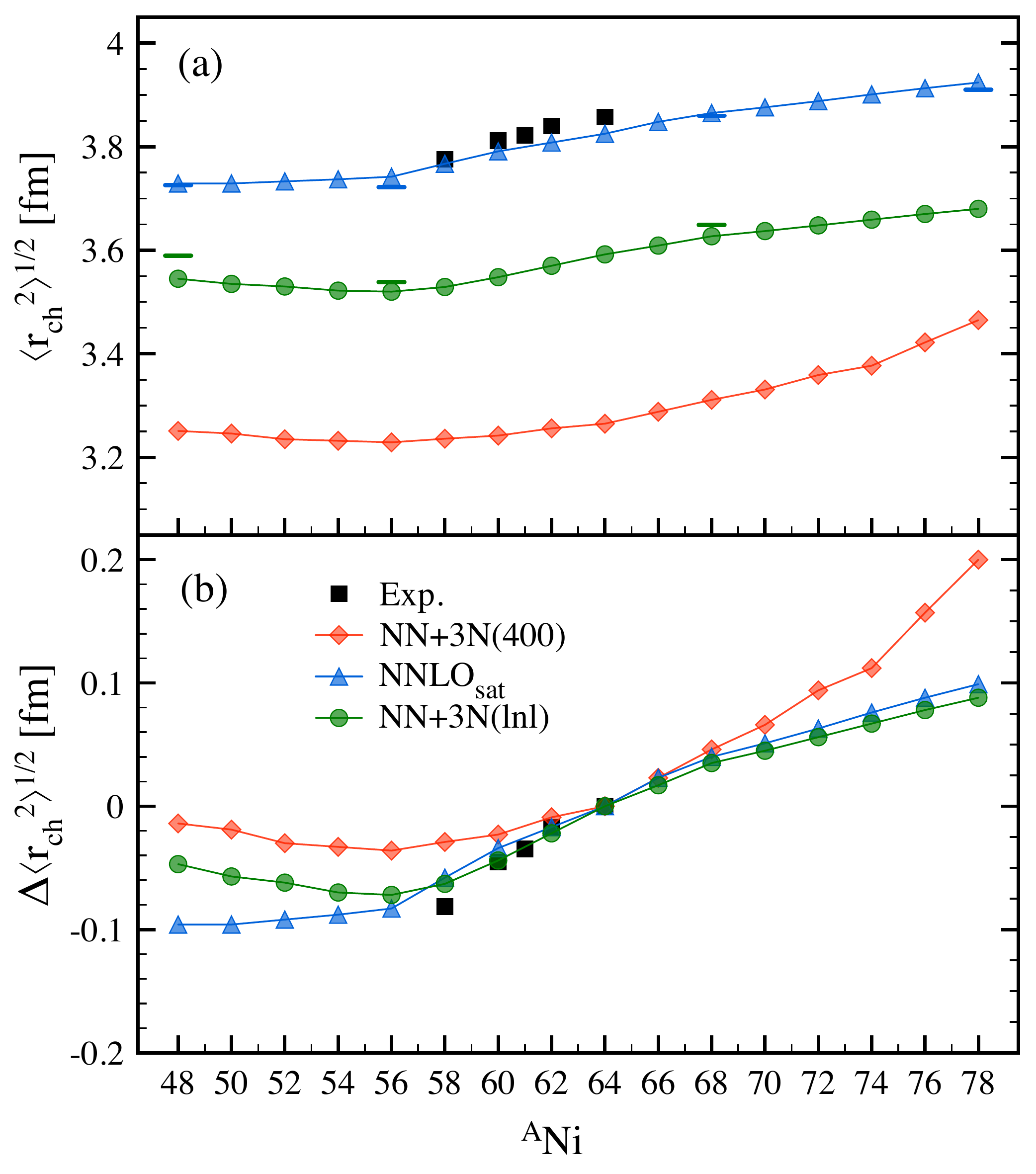}
\caption{Same as Fig.~\ref{fig_radii_O} but for nickel isotopes. In panel (b), shifts are relative to $^{64}$Ni. Experimental data are taken from Refs.~\cite{Angeli13}.
The estimated computational errors due to model space truncations are below 2.5\% of the charge radius for \sat{}, while calculations are substantially converged for \lnl{} and  \ema{} (see Fig.~\ref{fig_NmaxHW}). 
The corrections to radii due to many-body truncation beyond ADC(3) are larger for \lnl{} and for \sat{} and remain $<$0.01~fm in all cases (see Fig.~\ref{fig_ADC_convergence}).}
\label{fig_radii_Ni}
\end{figure}

Next let us examine rms charge radii $\langle r^2_\text{ch} \rangle^{1/2}$. 
In the present approach rms charge radii are computed from rms point-proton radii by correcting for the finite charge distributions of protons and neutrons, as well as for the Darwin-Foldy term (see Ref.~\cite{Cipollone15} for details). 
ADC(2) results for absolute rms charge radii of oxygen, calcium and nickel isotopes are displayed in Figs.~\ref{fig_radii_O}(a), \ref{fig_radii_Ca}(a) and \ref{fig_radii_Ni}(a), all compared to available experimental data. 
Relative rms radii $\Delta \langle r^2_\text{ch} \rangle^{1/2}$, i.e. charge radii differences\footnote{Note that the relative rms radii used here, differences of $\langle r^2_\text{ch} \rangle^{1/2}$, differ from the \textit{mean square} shifts $\delta\langle r^2 \rangle$ sometimes found in the literature, defined as differences of $\langle r^2_\text{ch} \rangle$.} relative to a reference isotope, are shown in the corresponding panels (b).
Following the analysis of Secs.~\ref{sec_MSconv} and~\ref{sec_MBconv}, conservative errors for \sat{}~(\lnl{} and \ema{}) are estimated to amount up to 1.8\%~(1,5\%) of the charge radius up to the calcium isotopes and of 2.6\%~(1\%) for the nickels, including $^{78}$Ni. 
Note that these errors are dominated by model space convergence in the case of \sat{} and by the many-body truncations for \lnl{} and \ema{}.

For what concerns absolute radii, a large variation between the different interactions is observed in all cases.
For oxygen isotopes, studies with \ema{} and \sat{} exist in the literature~\cite{Lapoux16, Cipollone15}, where it was shown that already in these light systems \ema{} leads to a strong underestimation of the size of nuclei.
From Fig.~\ref{fig_radii_O}(a) one notices that the new \lnl{} interaction improves on \ema{} results producing charge radii that are $\sim$0.1 fm larger, reducing by a factor 2 or better the discrepancy with experiment.
A similar picture emerges from the analysis of calcium and nickel chains.
In calcium isotopes, see Fig.~\ref{fig_radii_Ca}(a), \ema{} strongly underestimates measured radii, with discrepancies of about 12-15$\%$ along the whole chain. \lnl{} significantly improves on \ema{} results producing charge radii that are $\sim$0.3 fm larger.  
Still, experimental data are underestimated by about 5-6$\%$ across all isotopes. 
\sat, on the other hand, succeeds in reproducing the bulk values of Ca rms radii, thus maintaining for this observable the good performances already observed for lighter nuclei~\cite{Duguet17a}.
Importantly, the present results with \sat{} are in good agreement with previous coupled-cluster calculations performed on closed-shell and neighbouring isotopes~\cite{Ruiz16}.
Similar conclusions can be inferred by inspecting results for nickel isotopes, reported in Fig.~\ref{fig_radii_Ni}(a). 
Absolute rms charge radii obtained with \sat{} maintain their good agreement with data even for this mass region.  
A kink is visible at $^{56}$Ni, in accordance with its good closed-shell character.
Beyond this point, the calculation follow the trend of the limited available data, slightly departing from experiment as neutron number increases.
Radii obtained with the other two interactions, again, severely underestimate experimental results. 
Discrepancies are in line with what observed in calcium isotopes, namely \ema{} is more than 15$\%$ off and \lnl{} about 8-9$\%$ off.
For all chains, ADC(3) calculations for closed-shell systems are also shown. 
Contrarily to total ground-state energies, radii are essentially converged at the ADC(2) truncation level and additional ADC(3) correlations do not change the overall picture.
In particular, it is clear that the discrepancies with experiment cannot be removed by improving the many-body truncation. 

When comparing radii obtained with SRG-evolved and bare Hamiltonians, an important caveat relates to the present omission of potentially relevant higher-body radius operators induced by the SRG transformation. 
In some recent calculations~\cite{Schuster2014OpEvol,Miyagi19}, such induced operators have been properly included for the \ema{} interaction but have not led to any sizeable improvement.
Thus, although it remains to be seen whether the same holds for \lnl{}, the discrepancies with the experimental data point to intrinsic deficiencies of the Hamiltonian that will have to be addressed in the future.

The systematic flaws in the underestimation of radii appear to be uniform across each isotopic chain. Thus, one may expect that they cancel out to a good extent in differential quantities putting in evidence the isospin dependence for each interaction. 
In Fig.~\ref{fig_radii_O}(b) rms radii differences relative to $^{16}$O are shown for oxygen.
Clearly, the spread of results is appreciably reduced, with small discrepancies showing up for the less stable systems.
Radii along the calcium chain, relative to $^{48}$Ca, are shown in Fig.~\ref{fig_radii_Ca}(b).
The lightest isotopes  $^{34-40}$Ca are the most sensible to the employed interaction, with rather different trends.
Interestingly, recent measurements in $^{36-38}$Ca~\cite{Miller19} appear to be in better agreement with \lnl{} results rather than the ones obtained with \sat{}, which predicts a somewhat steeper slope.
Both interactions based on the N$^{3}$LO two-nucleon force of Ref.~\cite{Entem03} predict an inversion of this trend when going down to mass $A$=34 while \sat{} does the opposite.
All calculations roughly reproduce the fact that charge radii for $^{40}$Ca and  $^{48}$Ca are basically the same.
However, none of them is capable of accounting for the parabolic behaviour between these two isotopes.
This is not surprising since this feature has been associated, in the contexts of particle-vibrations coupling and shell model calculations~\cite{Barranco1985CaRad,Caurier01}, to the presence of highly collective many particle-many hole configurations that are missing in the many-body approach employed here.
Energy density functionals are also striving to reproduce the experimental trend, with only recent applications based on Fayans functionals~\cite{Reinhard17} able to capture the peculiar behaviour.
After $^{48}$Ca, \sat{} and \lnl{} do improve on the poor trend of \ema{} but still fail to reproduce quantitatively the steep slope leading to $^{52}$Ca.
The charge radius of the latter, recently measured in laser spectroscopy experiments~\cite{Ruiz16}, thus remains a challenge for many-body calculations.
Relative radii in nickel isotopes can be examined in Fig.~\ref{fig_radii_Ni}(b). 
Here two distinct regions can be identified.
Below $^{58}$Ni, we find a similar behaviour to the one of Ca: \ema{} and \lnl{} follow the similar trends and actually predict an increase of rms radii with decreasing neutron number, while \sat{} does the opposite and decreases towards $^{48}$Ni. More experimental data on both proton rich Ca and Ni would be very useful to pin down this effect.
Above $^{58}$Ni, \lnl{} and \sat{} predict a very similar behaviour, while \ema{} shows a rather steep increase all the way up to $^{78}$Ni.
The limited amount of available data gives a stronger support to the former trend.
Also in this respect, an extension of our experimental knowledge to some of the neutron-rich nickel isotopes would be very valuable.
\begin{figure}[t]
\centering
\includegraphics[width=7.3cm]{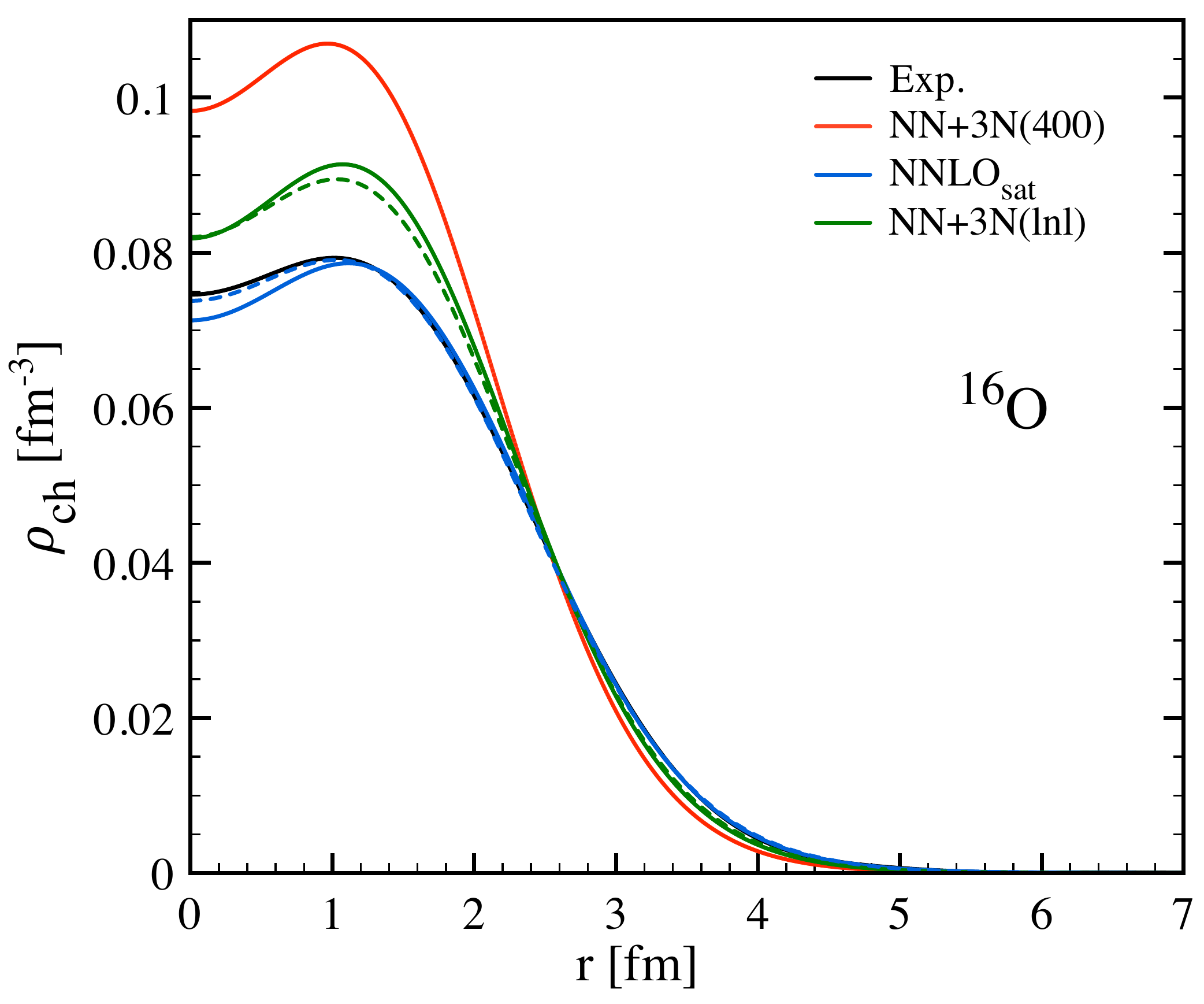}
\caption{Charge density distribution of $^{16}$O computed in ADC(2) (solid lines) and ADC(3) (dashed lines) with the \ema, \lnl{} and \sat{} interactions, together with the experimental distribution~\cite{deVries87}.}
\label{fig_rho_O}
\end{figure}
\begin{figure}
\centering
\includegraphics[width=7.3cm]{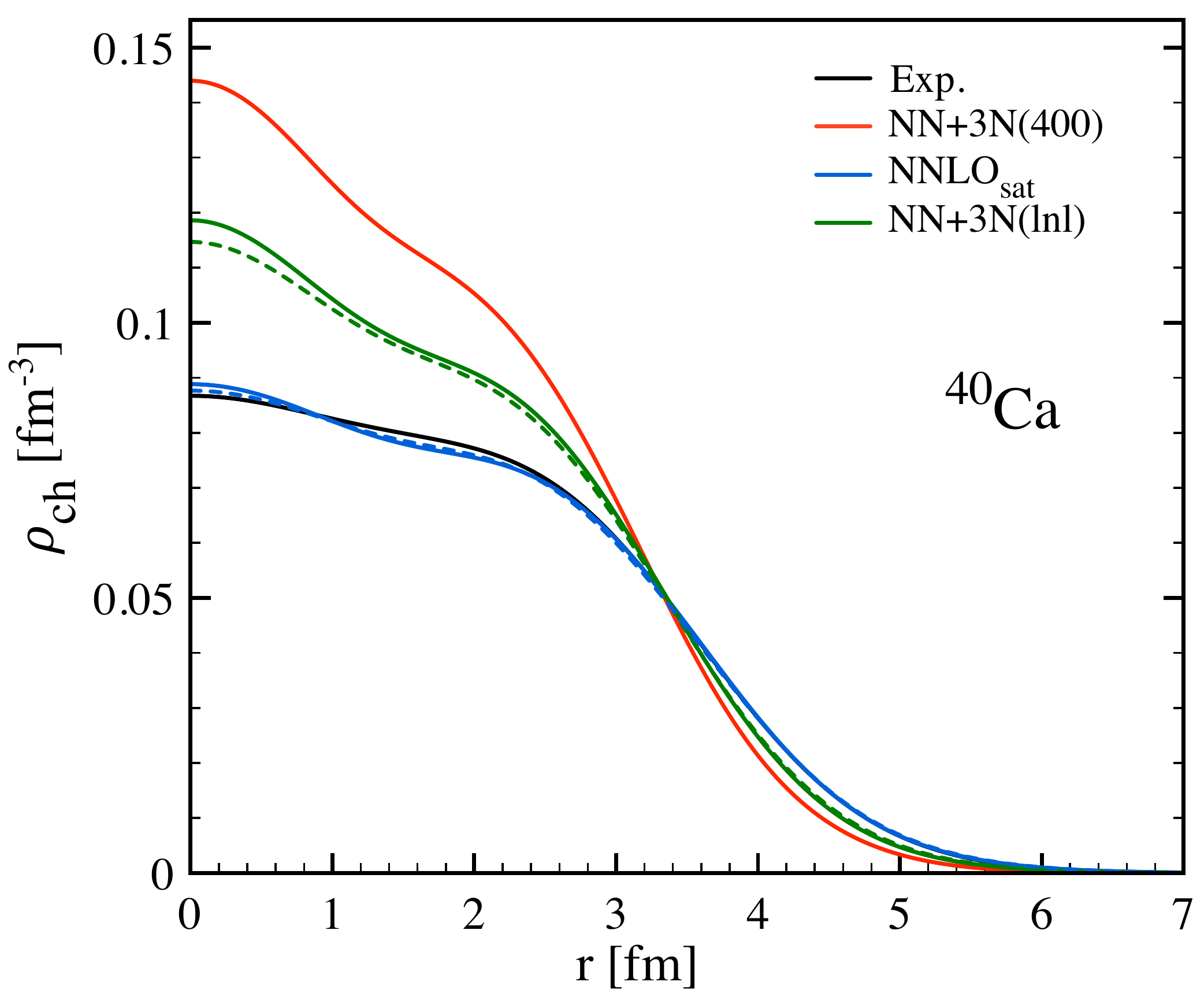}
\caption{Same as Fig.~\ref{fig_rho_O} but for $^{40}$Ca. Experimental data are taken from Refs.~\cite{deVries87, Emrich83}.}
\label{fig_rho_Ca}
\end{figure}

\begin{figure}
\centering
\includegraphics[width=7.3cm]{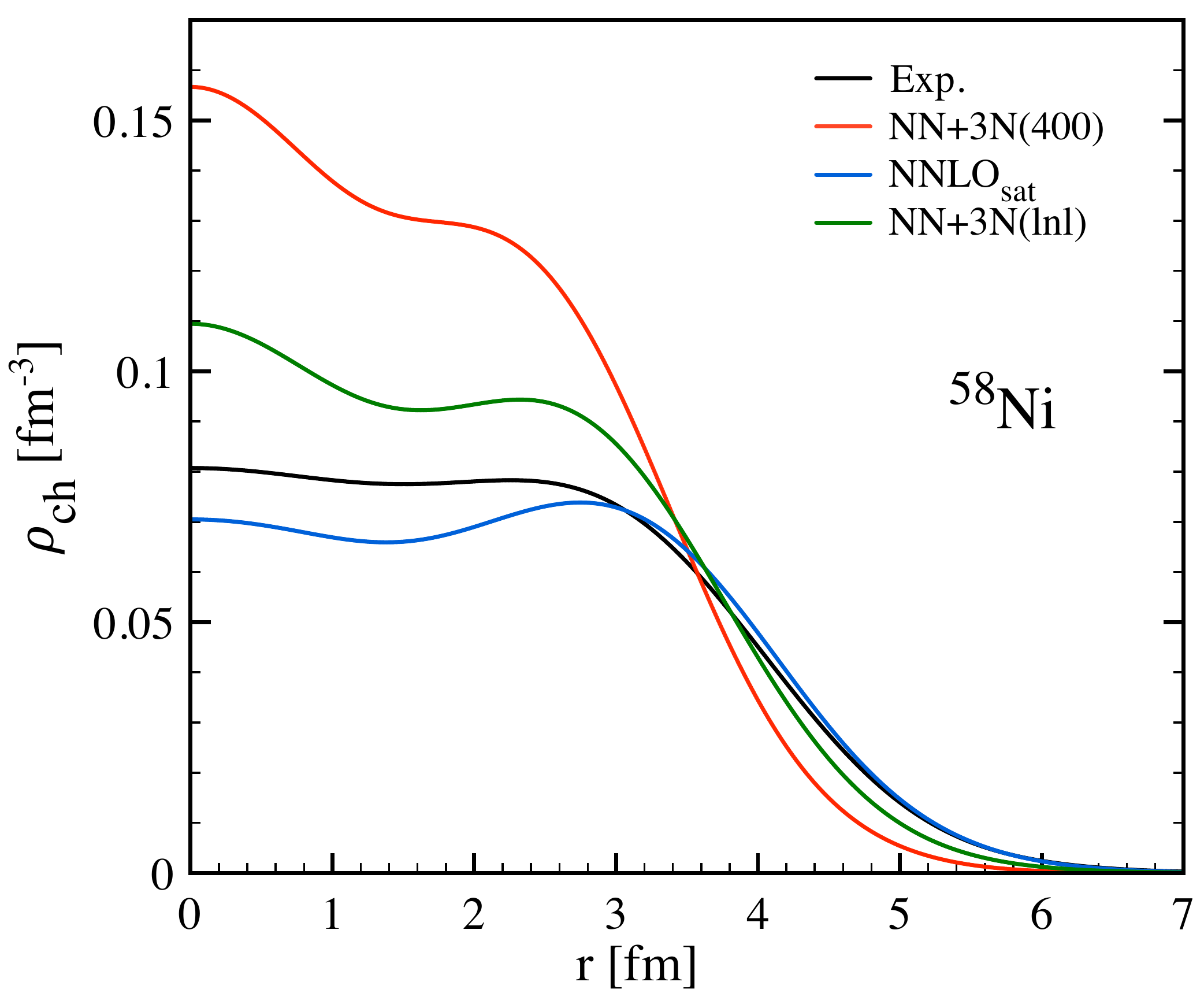}
\caption{Same as Fig.~\ref{fig_rho_O} but for $^{58}$Ni. Experimental data are taken from Ref.~\cite{deVries87}.}
\label{fig_rho_Ni}
\end{figure}

An even more stringent test for many-body calculations is represented by the description of the charge density distributions, from which rms charge radii have been traditionally extracted.
Charge distributions are experimentally accessed via electron scattering measurements and thus currently limited to stable nuclei, although considerable progress is being made towards extending this technique to unstable systems~\cite{Suda17}.
In the present framework the nuclear charge density is computed through the folding of the nuclear point-proton density distribution with the charge density distribution of the proton, see Ref.~\cite{Duguet17a} for details.
One representative isotope for each of the three chains is studied, namely $^{16}$O, $^{40}$Ca and $^{58}$Ni, respectively displayed in Figs.~\ref{fig_rho_O}, \ref{fig_rho_Ca} and \ref{fig_rho_Ni}. 
In all cases, unsurprisingly, \sat{} calculations are the closest to experiment.
For $^{16}$O and $^{40}$Ca the agreement is remarkable.
The description slightly deteriorates for $^{58}$Ni where, in spite of an excellent reproduction of the experimental charge radius, the theoretical result mildly deviates from the measured distribution.
As for charge radii, \lnl{} calculations largely improve on the poor performance of \ema{} but fail to reach the accuracy achieved by \sat{}.
ADC(3) calculations are also reported (dashed lines) in the case of $^{16}$O and $^{40}$Ca.
As for radii, one concludes that density distributions are largely converged at the ADC(2) level, although ADC(3) correlations do provide a refined description, e.g. in the central region of $^{16}$O.

\subsection{Excitation spectra}
\label{sec_spectra}

\begin{figure*}[th]
\centering
\includegraphics[width=\textwidth]{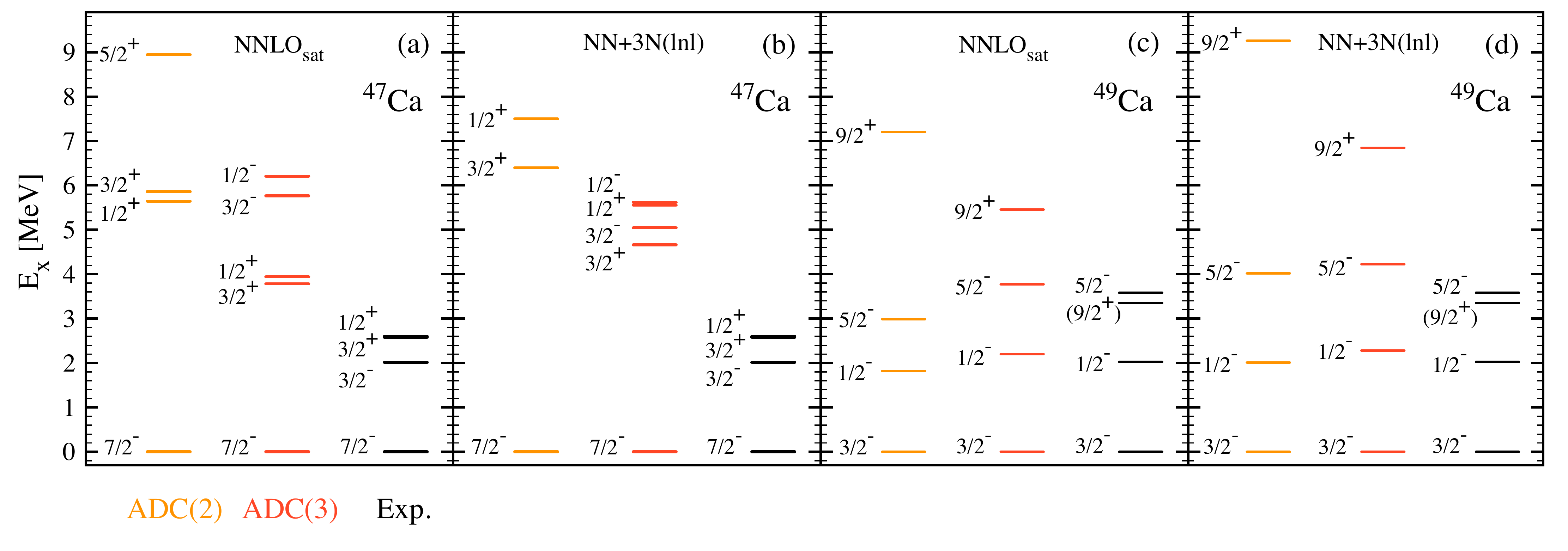}
\caption{One-neutron removal [panels (a) and (b)] and addition [panels (c) and (d)] spectra from and to $^{48}$Ca.
ADC(2) and ADC(3) calculations with \sat{} and \lnl{} interactions are compared to available experimental data~\cite{BurrowsA47, BurrowsA49}.
Computed low-lying states with spectroscopic factor larger than 1$\%$ and corresponding experimental energies are displayed.}
\label{fig_spectra_ADC}
\end{figure*}
\begin{figure*}
\centering
\includegraphics[width=\textwidth]{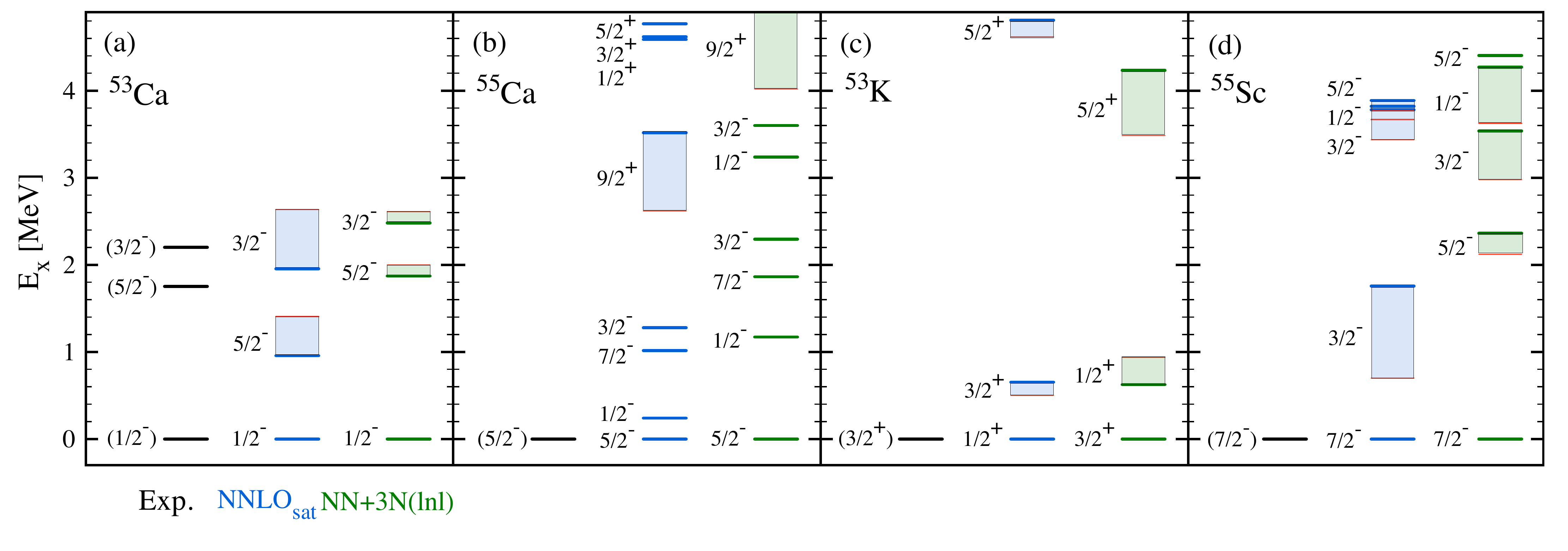}
\caption{One-nucleon addition and removal spectra from and to $^{54}$Ca.
Available experimental values are displayed as black thick lines, in the left column of each panel.
ADC(2) calculations are represented by blue (second column) and green (third column) thick lines for \sat{} and \lnl{} respectively. 
The red thin lines are ADC(3) energies for both Hamiltonians, with shaded areas connecting the corresponding  ADC(2) and ADC(3) values where available.
For $^{55}$Ca, states obtained via one-neutron removal from $^{56}$Ca in the ADC(2) approximation are also shown.
Low-lying states with $E_x<5$ MeV and spectroscopic factor larger than 10$\%$ are displayed.}
\label{fig_spectra_Ca}
\end{figure*}
Spectroscopic properties reflect some of the general features of a Hamiltonian (for instance the ability to reproduce magic gaps) but at the same time are sensitive to finer details, e.g. depending on the spin and parity of the excited state.
In Green's function theory, one-nucleon addition and removal (i.e., separation) energies are naturally accessed from the spectral representation of the one-body propagator, see Refs.~\cite{Dickhoff04, Barbieri17} for details.
The generalisation to Gorkov Green's functions allows for an analogous spectral form that also contains information on separation energy spectra of odd-even neighbours~\cite{Soma11a}.
While the ADC(2) approximation does introduce dynamical correlations that induce a fragmentation of the mean-field spectral function, one might ask whether such correlations are too crude for a quantitative description of (low-lying) excitation spectra.
The ADC(3) truncation scheme, by coupling the bare two particle-one hole (2p1h) and 2h1p (or three-quasiparticle in Gorkov theory) configurations introduced in ADC(2), stabilises dominant quasiparticle peaks, usually compresses the spectra and generates further fragmentation~\cite{Cipollone15,vonNiessen1984QCadc}.
\begin{figure}[t]
\centering
\includegraphics[width=8.5cm]{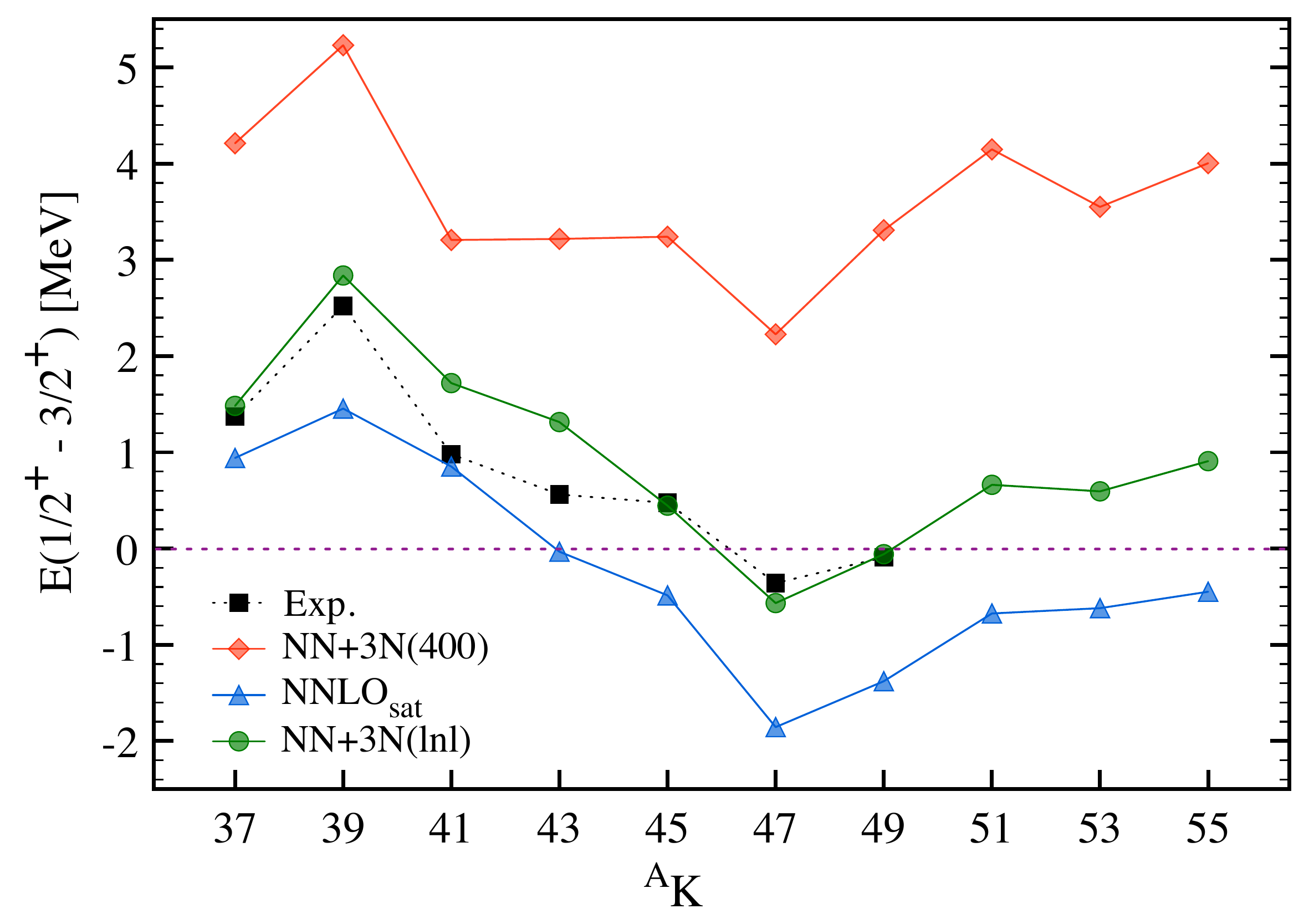}
\caption{Evolution of ground and first excited states along even-$N$ potassium isotopes. GSCGF ADC(2) calculations performed with three different interactions are compared to existing experimental data.}
\label{fig_K_spectra_exp}
\end{figure}

In order to test the two levels of approximation, one-neutron removal (addition) spectra from (to) $^{48}$Ca are studied in detail in Fig.~\ref{fig_spectra_ADC} using \sat{} and \lnl{} Hamiltonians.
Starting with one-neutron removal, i.e. states in $^{47}$Ca, one first notices that, for both interactions, ADC(2) spectra are too spread out, with the first excited states at 5-6 MeV to be compared with about 2~MeV in experiment.
Since such states are associated to the removal of a neutron in the $sd$-shell, this is a direct consequence of the overestimation of the $N=20$ gap, see Fig.~\ref{fig_BE_Ca}(b).
Note that the overestimation is more severe in \lnl{} calculations, which is reflected in higher excitation energies and a larger splitting between the $1/2^+$ and $3/2^+$ states as dictated by its underestimation of radii.
Including ADC(3) correlations helps in compressing the spectrum, although the effect of an overestimated $N=20$ gap remains.
Interestingly, in \sat{} calculations the correct ordering of $1/2^+$ and $3/2^+$ states is re-established.
In addition, negative-parity states $1/2^-$ and $3/2^-$ appear.
As opposed to the positive-parity levels that are obtained as a simple removal from the $sd$-shell, such states correspond to more complex configurations involving particle-hole excitations across the $N=28$ gap and are not captured by the simpler ADC(2) approximation.

The situation is different for one-neutron addition spectra, displayed in Figs.~\ref{fig_spectra_ADC}(c) and \ref{fig_spectra_ADC}(d).
Here low-lying states computed at the ADC(2) level are already in reasonably good agreement with experimental values for both interactions.
Again, the quality of the description is correlated with the (excellent) reproduction of the $N=28$ and $N=32$ gaps over which the excitations take place.
ADC(3) correlations mainly impact \sat{} results, with the position and splitting of $1/2^-$ and $5/2^-$ changing by a few hundreds keV and moving closer to data.
An exception is represented by the $9/2^+$ state, which is high-lying in ADC(2) and gets shifted down by a few MeV in ADC(3).

Following this analysis let us turn to the case of $^{54}$Ca and look at all one-nucleon removal and addition spectra, i.e. at its four possible odd-even neighbours.
These isotopes are of significant interest, with experiments that either have been recently performed or that are planned for the near future, and should complement the currently scarse data in the region.
Results from both ADC(2) and ADC(3) calculations with \sat{} and \lnl{} are reported in Fig.~\ref{fig_spectra_Ca}.
In $^{53}$Ca two excited states have been measured around 2 MeV with tentative spin-parity assignments of  $5/2^-$ and  $3/2^-$~\cite{Perrot2006betaK,Steppenbeck2013NatCa54}.
Both interactions yield the two states and support the spin assignments.
However, \lnl{} does a better job in reproducing both the position and the energy splitting between them.
In $^{55}$Ca, in addition to one-neutron addition states to $^{54}$Ca, one-neutron removal states from $^{56}$Ca in the ADC(2) approximation are also shown.
The spectra generated by the two interactions display the same low-lying states, although the one from \sat{} is more compressed than the one from \lnl.
In both cases the excited state corresponding to the main one-neutron addition quasiparticle, with spin-parity $9/2^+$, shows a large correction from ADC(3).
In $^{53}$K the two Hamiltonians predict a different ground state, with \lnl{} agreeing with the tentative experimental assignment.
Finally, in $^{55}$Sc it is the first excited states to be different, with \sat{} and \lnl{} predicting, respectively, $3/2^-$ and $5/2^-$ states on top of the $7/2^-$ ground state.

The identification of the ground-state spin in $^{53}$K is of particular interest and is related to a series of present-day experimental efforts along potassium isotopes.
From $^{37}$K up to $^{45}$K, the ground-state spins have been known to be $3/2^+$, as a naive shell model picture would suggest.
Several years back, $^{47}$K was shown to have a $1/2^+$ ground state via a laser spectroscopy experiment~\cite{Touchard82}, with $3/2^+$ becoming a low-lying excited state at 360~keV.
Recently, high-resolution collinear laser spectroscopy measurements determined that the ground-state spin inversion is maintained in $^{49}$K but a re-inversion occurs for $^{51}$K~\cite{Papuga13}.
Available experimental data are summarised in Fig.~\ref{fig_K_spectra_exp}, where the energy difference between $1/2^+$ and  $3/2^+$ states is displayed for even $N$ potassium isotopes.
At the time, GSCGF calculations were performed with the \ema{} interaction~\cite{Papuga14}, which resulted in the red curve  reported in Fig.~\ref{fig_K_spectra_exp}.
Although the calculations parallel the experimental trend, the energy gap between the two states is largely overestimated, and the spin inversion in $^{47}$K is absent.
The same observables have been computed here using the two more recent interactions.
\sat{} captures the trend as $N$ increases, but presents a shift compared to data that generates the inversion already at $^{43}$K. 
After that, the ground-state is predicted to have always spin-parity $1/2^+$.
Instead, \lnl{} succeeds in reproducing experimental data, including the inversion and re-inversion of the ground-state spin-parity and the position of the first excited state with remarkable accuracy.
Note that Fig.~\ref{fig_K_spectra_exp} displays results at the ADC(2) level. The ADC(3) corrections for the $^{53}$K gap, from Fig.~\ref{fig_spectra_Ca}, are of at most 0.4~MeV and suggest that missing many-body truncations could shift slightly these curves but are unlikely to alter our conclusions.

\section{Discussion and conclusions}
\label{sec_discussion}

\begin{figure}[t]
\centering
\includegraphics[width=8cm]{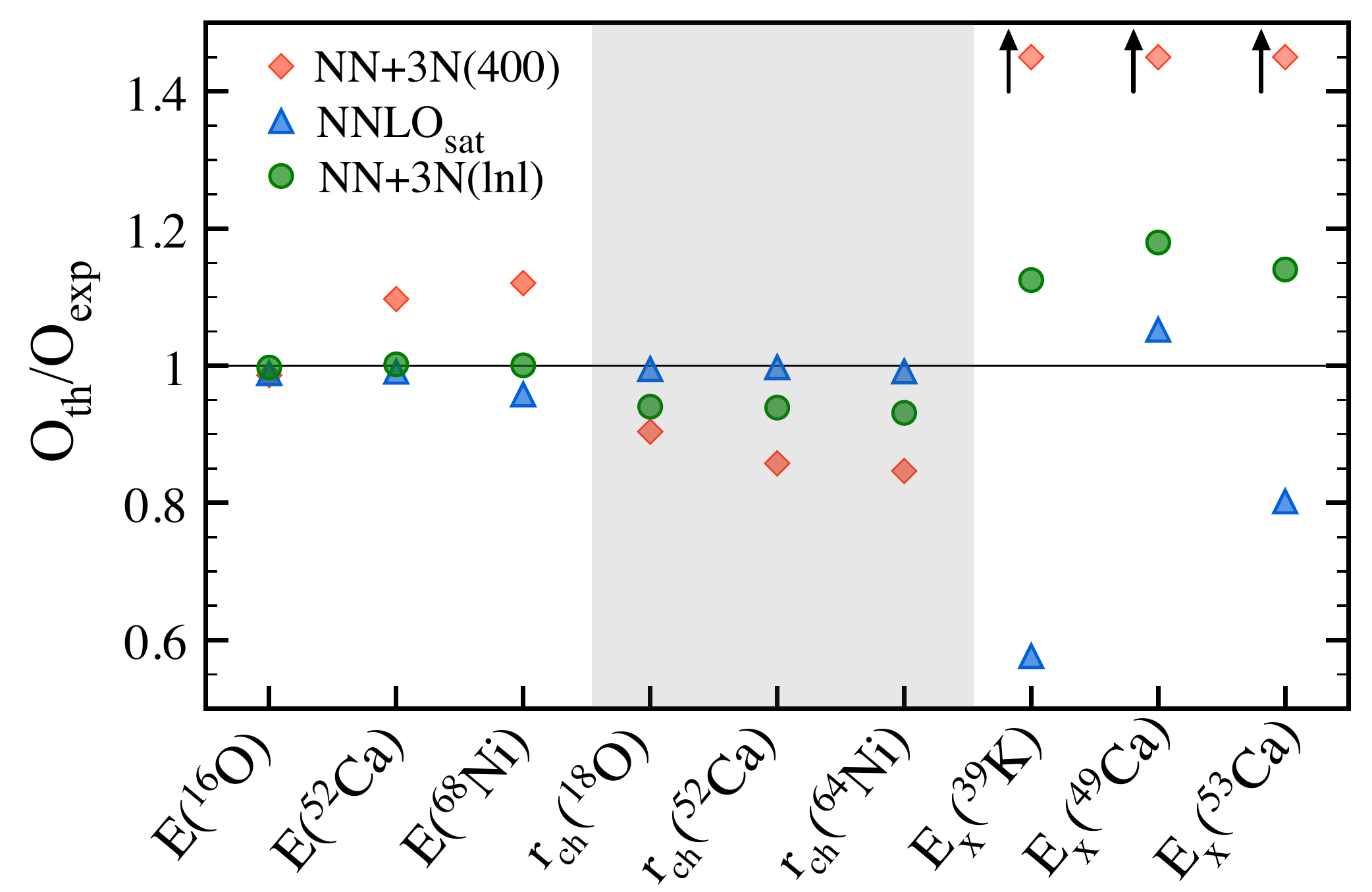}
\caption{Ratio of theoretical and experimental values for various observables computed with the \ema, \lnl{} and \sat{} interactions.
Binding energies and one-nucleon separation energies (rms charge radii) are evaluated at the ADC(3) [ADC(2)] level.
For \sat{}, radii are calculated at $\hbar \Omega=$~14~MeV, which represents the optimal value in terms of model-space convergence (see Fig.~\ref{fig_NmaxHW}).
All other observables are taken from the sets of calculations presented above.
Excitation energies computed with \ema{} lie outside the range shown in the figure.
For all observables and interactions, uncertainties coming from model-space and many-body truncations are of the order of the symbol sizes in the plot.
}
\label{fig_summary}
\end{figure}

Figure~\ref{fig_summary} summarises the performance of the three Hamiltonians on the different observables considered in the present work.
Representative ground-state energies, rms charge radii and one-nucleon separation energies are displayed. 
The older \ema{} interaction served as a workhorse in the early applications of \textit{ab initio} calculations with chiral $2N+3N$ forces, with empirical success in light isotopes, up to oxygen isotopes and neighbouring elements.
In particular, a notable achievement was the correct reproduction of the oxygen dripline~\cite{Otsuka10, Hergert13, Cipollone13}.
Typically employed in combination with SRG evolution, it has allowed important benchmarks between many-body methods, both non-perturbative and perturbative, that gave the practitioners confidence in the quality of the different many-body approximations~\cite{Hebeler15}.
However, its flaws appeared evident early on with strong overbinding being generated as mass number increases~\cite{Binder14} and a severe underestimation of nuclear radii even for oxygen isotopes~\cite{Lapoux16}.
Furthermore, formal issues were recently raised~\cite{Schiavilla18, Gazit19}, which question the consistency of the calculations that employed this interaction.
The overbinding and underestimation of radii emerging in \ema{} calculations are clearly visible in Fig.~\ref{fig_summary}.

Present results also confirm the overall empirical quality of the \sat{} interaction. 
Ground-state energies are well reproduced even beyond the light nuclei that were used in the fit of its coupling constants.
A mild underbinding is observed for heavier nickel isotopes.
However, before drawing definitive conclusions a careful study of model-space convergence (in terms of the one- and three-body truncations $e_{\text{max}}$ and $e_{3\text{max}}$, see also Figs.~\ref{fig_NmaxHW} and~\ref{fig_ramp}) has to be performed.
Nevertheless, contributions to charge radii are excellently reproduced by this interaction, even for nickel isotopes. 
The fact that radii, their associated density distributions and trends in the ground-state energies are already well converged at the ADC(2) opens the way to systematic calculations of full isotopic chains within the GSCGF approach.
Concerning spectroscopic properties, \sat{} had proven very accurate in the neutron $pf$-shell for $^{34}$Si and $^{36}$S~\cite{Duguet17a}.
This is to a good extent confirmed here in the neutron addition and removal spectra of $^{52}$Ca and $^{54}$Ca.
The agreement with experiment however deteriorates when looking at the $sd$-shell below $N=20$, with this Hamiltonian struggling to reproduce the observed inversion and re-inversion of the ground and first excited states along potassium isotopes.
\begin{figure}[t]
\centering
\includegraphics[width=8.6cm]{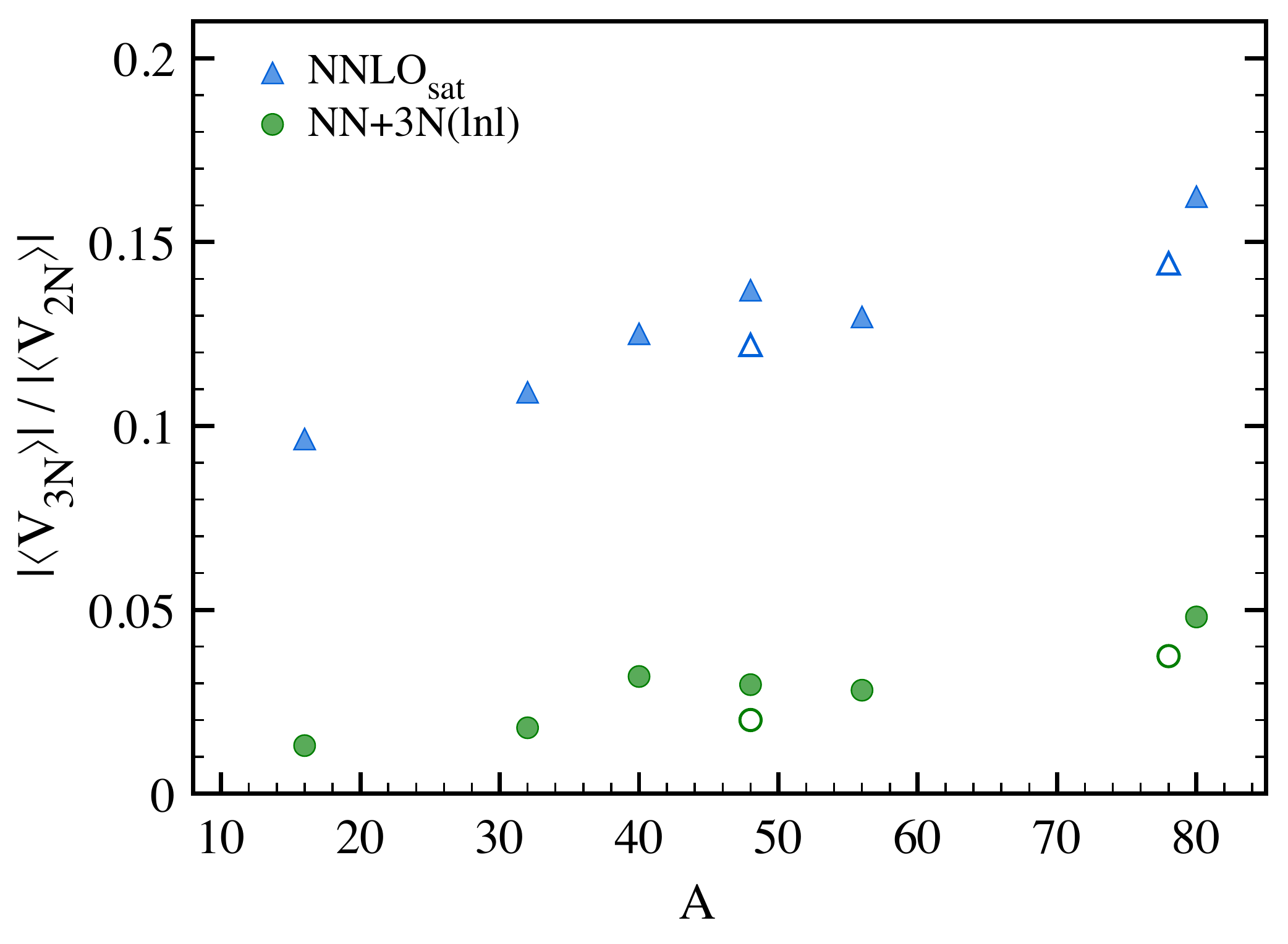}
\caption{Ratio of expectation values of three- ($V_{\text{3N}}$) and two-body ($V_{\text{2N}}$) operators in the \sat{} and \lnl{} Hamiltonians.
In $V_\text{2N}$, the two-body part of the centre-of-mass kinetic energy has been subtracted.
For the \lnl{} interaction, $V_{\text{3N}}$ contains original (i.e. SRG-unevolved) three-body forces while induced three-body operators have been included in $V_{\text{2N}}$.
Calculations are performed at the ADC(2) level.
Results are shown for $N=Z=\{ 8, 16, 20, 24, 28, 40\}$ nuclei (full symbols), plus $^{48}$S and $^{78}$Ni (empty symbols).}
\label{fig_2Nvs3N}
\end{figure}

So far, the novel \lnl{} interaction had been applied only to specific cases~\cite{Leistenschneider18, Gysbers19, Chen19}, but never tested in a systematic way.
In the present work its main ground-state properties as well as some selected excitation spectra have been studied extensively in light and medium-mass nuclei.
Results in light systems are very encouraging, with NCSM calculations in overall good agreement with experiment even for spectra that are known to be particularly sensitive to nuclear forces.
Total energies are well reproduced across the whole light sector of the nuclear chart.
In medium-mass nuclei, present calculations focused on three representative isotopic chains.
Total binding energies are found to be in remarkable agreement with experimental values all the way up to nickel isotopes once ADC(3) correlations are included, thus correcting for the overbinding generated with \ema.
ADC(2) calculations of differential quantities, where ADC(3) contributions essentially cancel out, are also very satisfactory and are able to capture main trends and magic gaps in two-neutron separation energies along all three chains.
As evidenced in Fig.~\ref{fig_summary}, although largely improving on \ema, rms charge radii obtained with the \lnl{} interaction still underestimate experimental results and do not reach the quality of \sat.
On the other hand this interaction yields an excellent spectroscopy, also where \sat{} strives to give even a qualitatively correct account of experimental data. 
One-nucleon addition and removal spectra in neutron-rich calcium are well reproduced.
Impressively, the evolution of the energy differences between the ground and first excited states along potassium isotopes follows closely the experimental measurements.

Further insight can be gained by gauging the importance of $3N$ operators in the two interactions.
In Fig.~\ref{fig_2Nvs3N} the ratio of $3N$ over $2N$ contributions to the total energy is displayed for a selection of nuclei as a function of mass number $A$ for \sat{} and \lnl.
In the former, $3N$ operators are much more relevant, reaching almost 20\% of the $2N$ contribution in heavier systems.
In contrast, the ratio stays rather low, around 5\%, for \lnl.
This has first of all practical consequences, as in the majority of many-body calculations the treatment of $3N$ operators is usually not exact, following either a normal-ordered two-body approximation (see, e.g., Ref.~\cite{Roth12}) or some generalisation of it~\cite{Carbone13}.
Hence a strong $3N$ component is in general not desirable.
On top of that, one might worry about the hierarchy of many-body forces from the standpoint of EFT, and possible need to include subleading $3N$ or $4N$ operators that could have a sizeable effect.

\begin{figure}
\centering
\includegraphics[width=8.6cm]{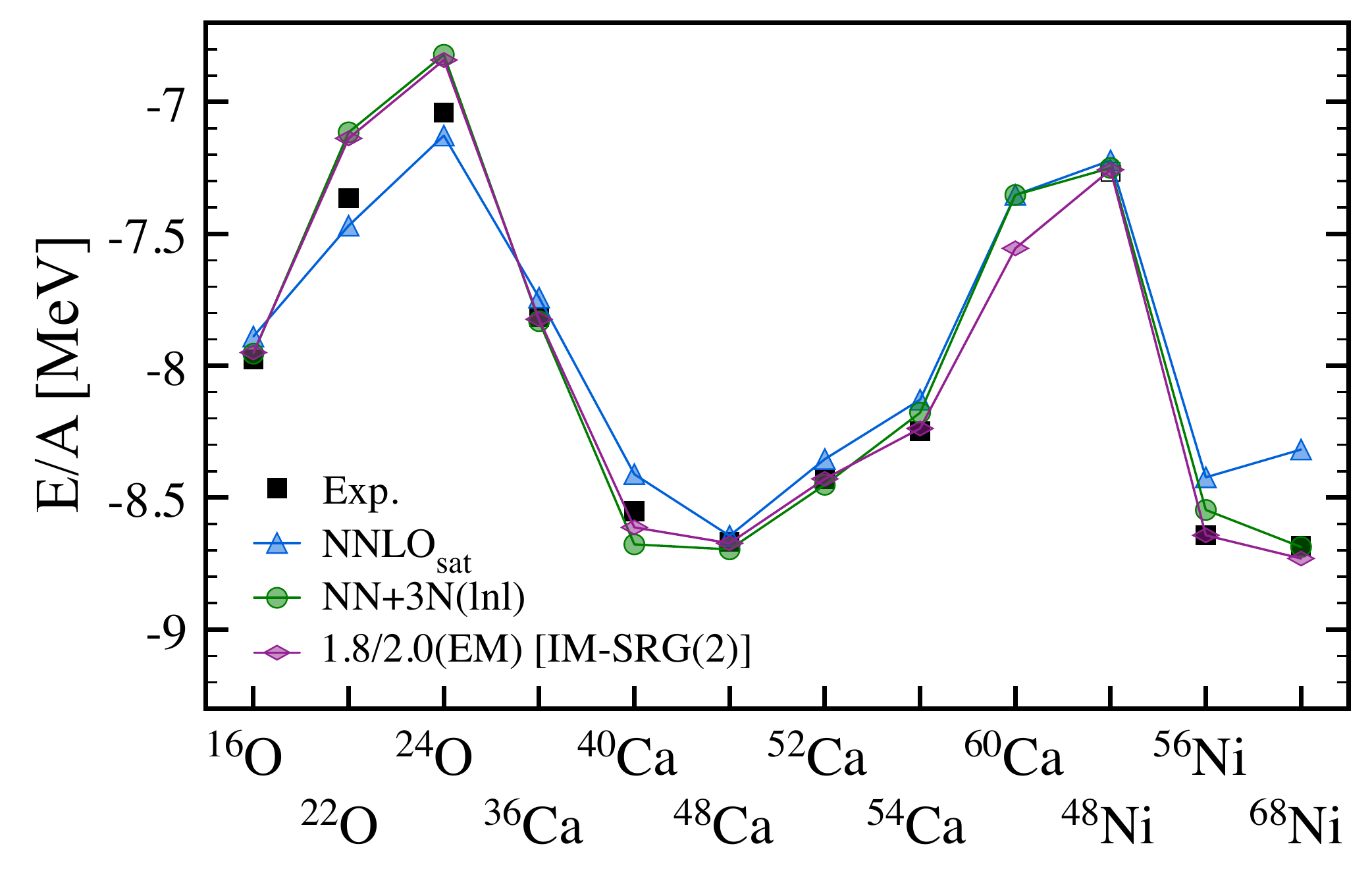}
\caption{Binding energy per particle for a set of doubly closed-shell nuclei computed with three different $NN+3N$ interactions and compared to available experimental data.
\sat{} and \lnl{} values come from the present work and refer to ADC(3) calculations.
\heb{} results were obtained via full-space IM-SRG(2) calculations and originally published in Ref.~\cite{Simonis17}.}
\label{fig_1820}
\end{figure}

Finally, let us compare \lnl{} and \sat{} to an interaction that has been extensively employed in nuclear structure studies in the past few years. 
Usually labelled as \heb{} and first introduced in Ref.~\cite{Hebeler11}, it has proven to yield an accurate reproduction of ground-state energies (as well as low-energy excitation spectra) over a wide range of nuclei~\cite{Simonis16,Simonis17,Gysbers19,Holt19}.
Furthermore, it leads to a satisfactory description of infinite nuclear matter properties~\cite{Hebeler11,Drischler14,Drischler19}.
In Fig.~\ref{fig_1820} binding energies per particle obtained within in-medium similarity renormalisation group (IM-SRG) calculations with the \heb{} interaction~\cite{Simonis17} are compared, for a set of closed-shell systems, to the ones computed at the ADC(3) level with \lnl{} and \sat{}.
The three sets of calculations achieve an overall excellent reproduction of experimental data.
While \sat{} is superior in light nuclei, it tends to slightly underbind some of the heavier systems.
One also notices a striking resemblance of the results obtained with the \lnl{} and \heb{} interactions (with the only exception of $^{60}$Ca, for which no experimental measurement exists) all the way up to $^{68}$Ni.
The two potentials indeed present several similarities.
First, the bare $NN$ part is the same~\cite{Entem03}, even though \lnl{} and \heb{} are subsequently SRG-evolved to different scales, $\lambda{=}2$~fm$^{-1}$ and $\lambda{=}1.8$~fm$^{-1}$ respectively.
Second, the $3N$ part builds on N$^2$LO operators and, in the case of \heb, non-local regulators are applied.
A difference comes from the fact that for \heb{} $3N$ forces are not SRG-evolved consistently with the $NN$ operators, but rather the LECs of the three-body contact terms are re-fitted \emph{a posteriori} to the energy of $^3$H and the charge radius of $^4$He.
In the end, this results in values ($c_D=1.264$ and $c_E=-0.120$) that are not very different from the ones of \lnl{} (see Table~\ref{tabLEC}).

The present systematic analysis shows that the novel \lnl{} Hamiltonian represents a promising alternative to existing nuclear interactions. 
In particular, it has the favourable features of (i) being adjusted solely on $A=2,3,4$ systems, thus complying with the \textit{ab initio} strategy, (ii) yielding an excellent reproduction of experimental energies all the way from light to medium-heavy nuclei and (iii) behaving well under similarity renormalisation group transformations, with small induced four-nucleon forces, thus allowing calculations up to medium-heavy systems with moderate computational costs.
A first large-scale application with SCGF calculations along few isotopic chains around $Z=20$ is already underway and confirms its excellent phenomenological properties~\cite{Soma19prep}.
In the short term, having such high-quality interactions at hand allows us to make useful predictions and to test in depth existing and forthcoming many-body methods.
In the long term, such efforts aim to contribute to the long-standing goal of performing simulations of atomic nuclei with fully controlled theoretical uncertainties.
\vspace{0.5cm}

\section*{Acknowledgements}

We thank the authors of Ref.~\cite{Simonis17} for providing the data included in Fig.~\ref{fig_1820}.
This work was supported by the NSERC Grant No. SAPIN-2016-00033,
by the Espace de Structure et de r\'eactions Nucl\'eaires Th\'eorique (ESNT) at CEA
and by the United Kingdom Science and Technology Facilities Council (STFC) under Grants No. ST/P005314/1 and No. ST/L005816/1.
Computing support for P.N. came from an INCITE Award on the Titan supercomputer of the Oak Ridge Leadership Computing Facility (OLCF) at ORNL and from Westgrid and Compute Canada. SCGF calculations were performed by using HPC resources from GENCI-TGCC (Contracts No. A003057392 and A005057392) and at the DiRAC  the DiRAC DiAL system at the University of Leicester, UK, (BIS National E-infrastructure Capital Grant No. ST/K000373/1 and STFC Grant No. ST/K0003259/1).
TRIUMF receives federal funding via a contribution agreement with the National Research Council of Canada. 

\bibliography{biblio}

\end{document}